\pdfoutput=1
\documentclass[aps,pre,twocolumn,showpacs,superscriptaddress,amsmath,amssymb,final,letterpaper]{revtex4}
\usepackage{graphicx}
\usepackage{subfigure}

\DeclareMathOperator{\Li}{Li}

\begin{document}
\title{Time evolution of epidemic disease on finite and infinite networks}
\author{Pierre-Andr{\'e} No\"el}
\affiliation{University of British Columbia Centre for Disease Control, Vancouver (British Columbia), Canada V5Z 4R4}
\affiliation{D\'epartement de Physique, de G\'enie Physique, et d'Optique, Universit\'e Laval, Qu\'ebec (Qu{\'e}bec), Canada G1V 0A6}
\author{Bahman Davoudi}
\affiliation{University of British Columbia Centre for Disease Control, Vancouver (British Columbia), Canada V5Z 4R4}
\author{Robert C. Brunham}
\affiliation{University of British Columbia Centre for Disease Control, Vancouver (British Columbia), Canada V5Z 4R4}
\author{Louis J. Dub\'e}
\altaffiliation{Also at Laboratoire de Chimie-Physique-Mati\`ere et Rayonnement, Universit\'e Pierre et Marie Curie, 75231 Paris 05, France}
\affiliation{D\'epartement de Physique, de G\'enie Physique, et d'Optique, Universit\'e Laval, Qu\'ebec (Qu{\'e}bec), Canada G1V 0A6}
\author{Babak Pourbohloul}
\email[Corresponding author:]{babak.pourbohloul@bccdc.ca}
\affiliation{University of British Columbia Centre for Disease Control, Vancouver (British Columbia), Canada V5Z 4R4}
\affiliation{Department of Health Care and Epidemiology, University of Bristish Columbia, Vancouver (British Columbia), Canada V6T 1ZA}
\date{\today}
\begin{abstract}
Mathematical models of infectious diseases, which are in principle analytically tractable, use two general approaches. The first approach, generally known as compartmental modeling, addresses the time evolution of disease propagation at the expense of simplifying the pattern of transmission. The second approach uses network theory to incorporate detailed information pertaining to the underlying contact structure among individuals while disregarding the progression of time during outbreaks. So far, the only alternative that enables the integration of both aspects of disease propagation simultaneously \emph{while preserving the variety of outcomes} has been to abandon the analytical approach and rely on computer simulations. We offer a new analytical framework, which incorporates both the complexity of contact network structure and the time progression of disease spread. Furthermore, we demonstrate that this framework is equally effective on finite- and ``infinite''-size networks. This formalism can be equally applied to similar percolation phenomena on networks in other areas of science and technology.
\end{abstract}
\pacs{89.75.Hc, 87.23.Ge, 05.70.Fh, 64.60.Ak}
\maketitle
%\tableofcontents
\section{Introduction}
The spread of communicable diseases is a dynamical process and as such, understanding and controlling infectious disease outbreaks and epidemics is pertinent to the temporal evolution of disease propagation. Historically, this aspect of disease transmission has been studied using coarse-grained dynamical representation of populations, known as compartmental models \cite{andersonmay91,brauer01,diekmann00,earn00_science,hethcote00_siamrev}. In these models, a population is divided into a number of epidemiological states (or classes) and the time evolution of each is described by a differential equation.

Although this approach, and its more complex variants, has been instrumental in understanding several features of infectious diseases over the past 3 decades, it comes with a major simplification. The simplifying assumption states that the population is ``well mixed'', \emph{i.e.}, every infectious individual has an equal opportunity to infect others. This assumption may be valid in the broader context of population biology. Human populations, however, tend to contact each other in a heterogeneous manner based on one's age, profession, socio-economic status or behavior, and thus, the well-mixed approximation cannot portray an accurate image of disease spread among humans \cite{meyers05_jtb}.

Recent advances in network- and percolation-theories, have paved the way for physicists to bring a new perspective to understanding disease spread. Over the past decade, seminal works by Watts and Strogatz on small-world networks \cite{watts98_nature,watts99}, Barabasi \emph{et al.} on scale-free networks \cite{albert02_rmp} and Dorogovtsev, Mendes \cite{dorogovtsev03}, Pastor-Satorras and Vespignani \cite{pastor-satorras04}, among others, on the dynamics of networks, have shed light on a number of intriguing aspects of epidemiological processes. In particular, groundbreaking work by Newman \emph{et al.} \cite{newman01_pre,newman02_pre,newman03_siamrev} has provided a strong foundation for the formulation of epidemiological problems using tools developed by physicists.

Various dynamical processes that propagate from neighbour to neighbour on complex (natural or artificial) networks, \emph{e.g.}, a crawler (or spider) browsing the World Wide Web or rumors spreading in a population, reveals interesting similarities with the spread of epidemics in human population \cite{moreno04a_pre,moreno04b_pre}. In the present work, we specifically focus on disease propagation as the dynamical phenomenon and use the associated terminology. Our methodology is however quite general and can be applied \emph{mutatis mutandis} to other processes that manifest similar dynamical properties.

We are primarily interested in diseases where infected individuals are eventually removed from the dynamics of the system (\emph{i.e.}, infection is followed by naturally-acquired immunity or death), implying that the same person cannot be infected more than once. At any given time, we call an individual ``susceptible'' if he has never been exposed to the disease; ``exposed'' if he has acquired the infection but not currently able to pass on the disease to another person; ``infectious'' if he is currently able to transmit the infection to others; ``removed'' if he became immune or succumbed to death after acquiring the infection; and finally, ``infected'' if he has been exposed to the infectious agent at least once in the past, regardless of his current state (\emph{e.g.} exposed, infectious, removed).

Network analysis using the generating function formalism, developed by Newman \emph{et al.}, has proven to be a powerful tool when analyzing the spread of disease within networks \cite{newman02_pre}. Without directly addressing the question of ``when the transmission occurred?'', it provides reliable results on the final size of an outbreak/epidemic by addressing the question of ``whether transmission occurred?''. The first question, \emph{i.e.} the time evolution of the system, is presently beyond the formalism as originally derived.

Several researchers have recognized the importance of incorporating the notion of time into the generating function formalism that describes percolation dynamics on networks. In broaching this issue, many notable advances have been made. Numerous contributions (for a recent review, see \cite{boccaletti06_pr}) have used an approach closely related to compartmental models to assign a higher strength of infection to nodes of higher degree \cite{pastor-satorras01_prl,pastor-satorras01_pre,moreno02_epjb}. More recently, Marder \cite{marder07_pre} has calculated the probability distribution of outbreak sizes as a function of time for infinite-size networks, while Volz \cite{volz08_jmathbio} has addressed the finite-size effect by deriving estimates of the mean size of a large-scale epidemic, rather than the probability distribution. Despite these advances, one is required to develop a truly integrated analytical framework that simultaneously encompasses the time progression of disease, the network finite-size effects and the wide variety of possible outcomes. The major steps towards an integrative formalism are the principal achievement of this paper.

The outline of the paper is as follows. In Sec. \ref{section:InfiniteNetFormalism} we define the type of dynamics that will be studied and recall some of the tools used by Newman \cite{newman02_pre} for infinite networks, to which we add the concept of generations and phase-space representation. In Sec. \ref{section:discretetimefinitenetwork} we extend these tools for uses on finite-size networks and Sec. \ref{section:discussion} presents some results of our analysis, compares them with some existing
models, and discuss possible extensions of the formalism.   
 In Sec. \ref{section:conclusion} we give our conclusions and Appendices complete the analysis of 
Sec. \ref{section:discretetimefinitenetwork} and \ref{section:discussion}.
\section{Formalism for infinite networks \label{section:InfiniteNetFormalism}}
We map a system of $N$ individuals to a \emph{network} in which each individual is represented by a \emph{node} (or vertex) and the connection between each pair of individuals is represented by a \emph{link} (or edge). Two nodes are \emph{neighbours} if they are joined by a link. Contrary to compartmental approaches, a network representation of a system takes into account that each node does not have the same probability of interacting with every other node; in fact, one interacts only with its topological neighbours.  Each node has a \emph{degree} $k_i$ (number of neighbours) and the set $\{k_i\}$ (called \emph{degree sequence}) partially defines the network. Similarly, the set of probabilities $\{p_k\}$ that a random node has degree $k$ is the \emph{degree distribution}.

In many practical situations, the degree distribution is, together with the size $N$ of the network, the only available information on the network structure. We consider the ensemble of all possible networks obtained by drawing a degree sequence from the provided degree distribution and then, for every node $i \in \{1, 2, \ldots, N\}$, randomly connect each of its $k_i$ links to those of other nodes (no \emph{self-loops}) \cite{newman02_pre}. The quantities obtained through this paper are averages on this ensemble. We add to this maximum entropy definition the restriction that two nodes cannot share more than one link (no \emph{repeated links}). For sparse graphs --- in which the number of links scales linearly with the number of nodes --- the probability for such an event decreases as $1/N$ and can be neglected for large networks \cite{newman01_pre}.

To perform Monte-Carlo simulations of epidemic propagations on a network, one requires an explicit knowledge of that network structure. We have used the following method \cite{newman02_pre} to produce a network belonging to the ensemble described 
above:
\begin{enumerate}
\item[i.]  generate a random degree sequence $\{ k_i \}$ of length $N$ subjected to the degree distribution $\{ p_k\}$;
\item[ii.] make sure that $\sum_i k_i$ is an even number since a link is composed of 2 ``stubs'';
\item[iii.] for each $i$, produce a node with $k_i$ stubs;
\item[iv.] randomly choose a pair of unconnected stubs and connect them together. 
                Repeat until all unconnected stubs are exhausted;
\item[v.] test for the presence of self-loops and repeated links. 
              Remove the faulty stubs by randomly choosing a pair of connected stubs and rewire them to the former stubs. 
              Repeat until no self-loop and/or repeated links are found.
\end{enumerate}
%Clearly, not all the networks generated using this algorithm will have a giant component. These networks are part of the
% canonical ensemble we are interested in and, as such, they are kept for a proper averaging procedure.
%
\subsection{Basic generating functions \label{subsection:generatingfunctions}}
With knowledge of the degree sequence of the physical or social network of interest, we can obtain the corresponding degree distribution $\{p_k\}$. Following Newman \emph{et al.} \cite{newman01_pre} and Newman \cite{newman02_pre}, we define the probability generating function (\emph{pgf}) \cite{wilf94} for a random node as
\begin{align}
G_0(x) & = \sum_{k=0}^\infty p_k x^k \quad , \label{eq:G0}
\end{align}
respecting the normalization $G_0(1) = \sum_{k = 0}^\infty p_k = 1$. The average degree, $z_1$, can be easily obtained from
\begin{align}
z_1 & = \langle k \rangle = \sum_{k=0}^\infty k p_k = G_0'(1) \quad ,
\end{align}
where the prime denotes the derivative with respect to the argument. The probability, $q_k$, that $k$ nodes could be reached from the node we arrived at by following a random link (excluding this link from the count), can also be derived as
\begin{align}
G_1(x) & = \sum_{k = 0}^\infty q_k x^k = \frac{\sum_k (k+1)p_{k+1} x^k}{\sum_k (k+1) p_{k+1}} = \frac{1}{z_1} G_0'(x) \quad . \label{eq:G1}
\end{align}

While $G_0(x)$ and $G_1(x)$ contain information about the structure of the physical network linking nodes within the epidemiological system, they do not contain any information about the risk of disease transmission between two neighbouring nodes. However, with the additional knowledge of the transmissibility, $T$ --- the probability that an infectious node will infect one of its neighbors --- we can determine the probability of infecting $l$ out of $k$ neighbors as
\begin{align}
p_{l|k} & = \binom{k}{l} T^l (1-T)^{k - l} \quad . \label{eq:plk}
\end{align}
The \emph{pgf} for the number of infections directly caused by the initially infected node (``patient zero'') is then
\begin{align}
\sum_{l=0}^\infty \sum_{k=l}^\infty p_k \, p_{l|k} \, x^l & = \sum_{k=0}^\infty \sum_{l=0}^k p_k \binom{k}{l} T^l (1-T)^{k - l} x^l \nonumber \\ & = G_0\left(1 + (x-1)T\right) \quad . \label{eq:G01pxm1T}
\end{align}
Similarly, the probability distribution for the number of infections directly caused by a node reached by following a random link is generated by $G_1(1 + (x-1)T)$.

We can continue in the same vein and obtain informative results about the final state of the population in an infinite network, after the outbreak/epidemic has ended \cite{newman02_pre,meyers05_jtb}; however, this approach in itself does not yield any information about the duration of the epidemic, its speed of propagation or other time-related quantities.
\subsection{Generations \label{subsection:generations}}
To study the progression of the outbreak/epidemic over the network, we adopt an approach based on \emph{generations of infection}. We define \emph{generation} $0$ as the initial infected node of the outbreak/epidemic; nodes of generation $g$ are those who acquire the disease from a member of generation $g - 1$.

There is clearly a causality link among generations and we would expect nodes of higher generations to become infected, on average, later than those in generations closer to the initial infected node. In a future contribution, we will look more closely into the relationship between generations and continuous time evolution.

We now extend the generating function formalism to introduce a new \emph{pgf} for an arbitrary generation $g$
\begin{align}
G_g(x) & = \begin{cases}
           G_0(x) & \text{($g = 0$)} \\
           G_1(x) & \text{($g \ge 1$)}
           \end{cases} \quad .
\end{align}
As in Eq. \eqref{eq:G01pxm1T}, the \emph{pgf} for the number of nodes that acquire infection directly from a single node of generation $g$ is given by $G_g\left( 1 + (x-1)T \right)$. This assumption holds when the total number of infected nodes in the current and previous generations is small compared to the size, $N$, of the network. In such a case, the probability of infecting a node that is already infected is proportional to $1/N$; we assume that a node cannot be infected twice and the propagation of the disease follows a tree-like structure (without a closed loop). This condition is fulfilled in large networks either when there is no giant component or when we limit ourselves to the first few generations. Section \ref{section:discretetimefinitenetwork} removes these limitations to some extent.

From the properties of the \emph{pgf}'s, the expected number of secondary infections caused directly by an infected node in generation $g$ is given by
\begin{align}
\langle l_g \rangle & = \left. \frac{d G_g\left(1 + (x-1)T\right)}{dx} \right|_{x = 1} =
\begin{cases}
T G_0'(1) & \text{($g = 0$)} \\
T G_1'(1) & \text{($g \ge 1$)}
\end{cases} \quad .
\end{align}
Notice that $\langle l_g \rangle$ is identical for every generation, except the first one. It corresponds to a fundamental quantity in epidemiology, the \emph{basic reproductive number} \cite{andersonmay91,diekmann00}
\begin{align}
R_0 & = \langle l_1 \rangle = \langle l_2 \rangle = \ldots = T G_1'(1) = T \frac{z_2}{z_1} \quad ,
\end{align}
where $z_2 = G_0''(1)$ is the expected number of second neighbors for a randomly chosen node. $R_0 < 1$ implies that the expected number of infectious nodes decreases in consecutive generations, leading to the extinction of the disease. Conversely, $R_0 > 1$ implies that the expected number of infectious nodes increases and can potentially lead to an epidemic, which is a \emph{giant component} of occupied links in the language of percolation theory \cite{stauffer94}. It is worth noting that $R_0 > 1$ alone does not guarantee the occurrence of an epidemic; indeed, some realizations may have a number of new infections below the mean value $R_0$, and therefore lead to the extinction of the disease.
\subsection{Phase-space representation \label{subsection:phasespace}}
To proceed further, we define the quantity $\psi_{sm}^g$ as the probability of having $s$ infected nodes by the end of the $g$-th generation, of which $m$ became infected during the $g$-th generation. This probability is generated by
\begin{align}
\Psi_0^g(x,y) = \sum_{s,m} \psi_{sm}^g x^s y^m \quad . \label{eq:Psi0gdef}
\end{align}
Each element of the (triangular) matrix, $\psi_{sm}^g$, can be seen as a possible ``state of infection'' where the $s$ and $m$ dimensions provide information about the ``position'' (number of infected) and ``momentum'' (new infections) in the infection space, respectively.

We know from the previous sections that the probability distribution for the number of nodes that acquire infection directly from a single node of generation $g - 1$ is generated by $G_{g-1}\left( 1 + (x-1)T \right)$. Moreover, the \emph{pgf} for the sum of two independent quantities is given by the product of their \emph{pgf}'s \cite{wilf94, newman01_pre}. Therefore, the probability $P(m|s',m')$ that each state $(s',m')$ of generation $g - 1$ leads to $m$ new infections in generation $g$ is generated by
\begin{align}
\sum_m P(m|s',m') x^m & = \left[ G_{g-1}\left( 1 + (x-1)T \right) \right]^{m'} \quad .
\end{align}
Also, the state $(s',m')$ has probability $\psi_{s'm'}^{g-1}$ at generation $g-1$ and thus makes a contribution $\psi_{s'm'}^{g-1} P(m|s',m')$ to the state $(s = s'+m,m)$ at generation $g$. Hence, we obtain the recurrence relation
\begin{align}
\sum_{s',m'} \psi_{s'm'}^{g-1} x^{s'} \left[ G_{g-1}\left( 1 + (xy - 1)T \right) \right]^{m'} = \sum_{s,m} \psi_{sm}^g x^s y^m \label{eq:RecurrPsi}
\end{align}
with the initial condition $\psi_{sm}^0 = \delta_{s1}\delta_{m1}$ ($\delta_{ij}$ is the Kronecker delta). The states for which $m' = 0$ are absorbing states; the region above the main diagonal ($m' > s'$) is forbidden; and the main diagonal ($m' = s'$) is only accessible for the initial condition.

Finally, inserting Eq. \eqref{eq:RecurrPsi} into Eq. \eqref{eq:Psi0gdef} provides the (\emph{forward}) recurrence relation (for $g \ge 1$)
\begin{align}
\Psi_0^g(x,y) = \sum_{s',m'} \psi_{s'm'}^{g-1}\  x^{s'} \left[ G_{g-1}\left( 1 + (xy - 1)T \right) \right]^{m'} \label{eq:Psi0gInf}
\end{align}
with the initial condition $\Psi_0^0(x,y) = xy$ or $\psi_{sm}^0 = \delta_{s1} \delta_{m1}$, where $\delta_{ij}$ is the Kronecker delta. Equation \eqref{eq:Psi0gInf} implies further the recurrence
\begin{align}
\Psi_0^g(x,y) & = \Psi_0^{g-1}\left( x, G_{g-1}\left( 1 + (xy - 1)T \right) \right) \quad .
\end{align}
Note that $\Psi_0^g(x,1)$ generates the probabilities
\begin{align}
p_s^g = \sum_m \psi_{sm}^g \label{eq:ps}
\end{align}
that $s$ nodes are infected at generation $g$, independent of the number of new infections. Similarly, $\Psi_0^g(1,y)$ generates the probabilities
\begin{align}
p_m^g = \sum_s \psi_{sm}^g
\end{align}
that $m$ nodes are infected during generation $g$.

\begin{figure*}[htb]
\mbox{
\subfigure[~$g = 2$]{\includegraphics[width = .47\linewidth]{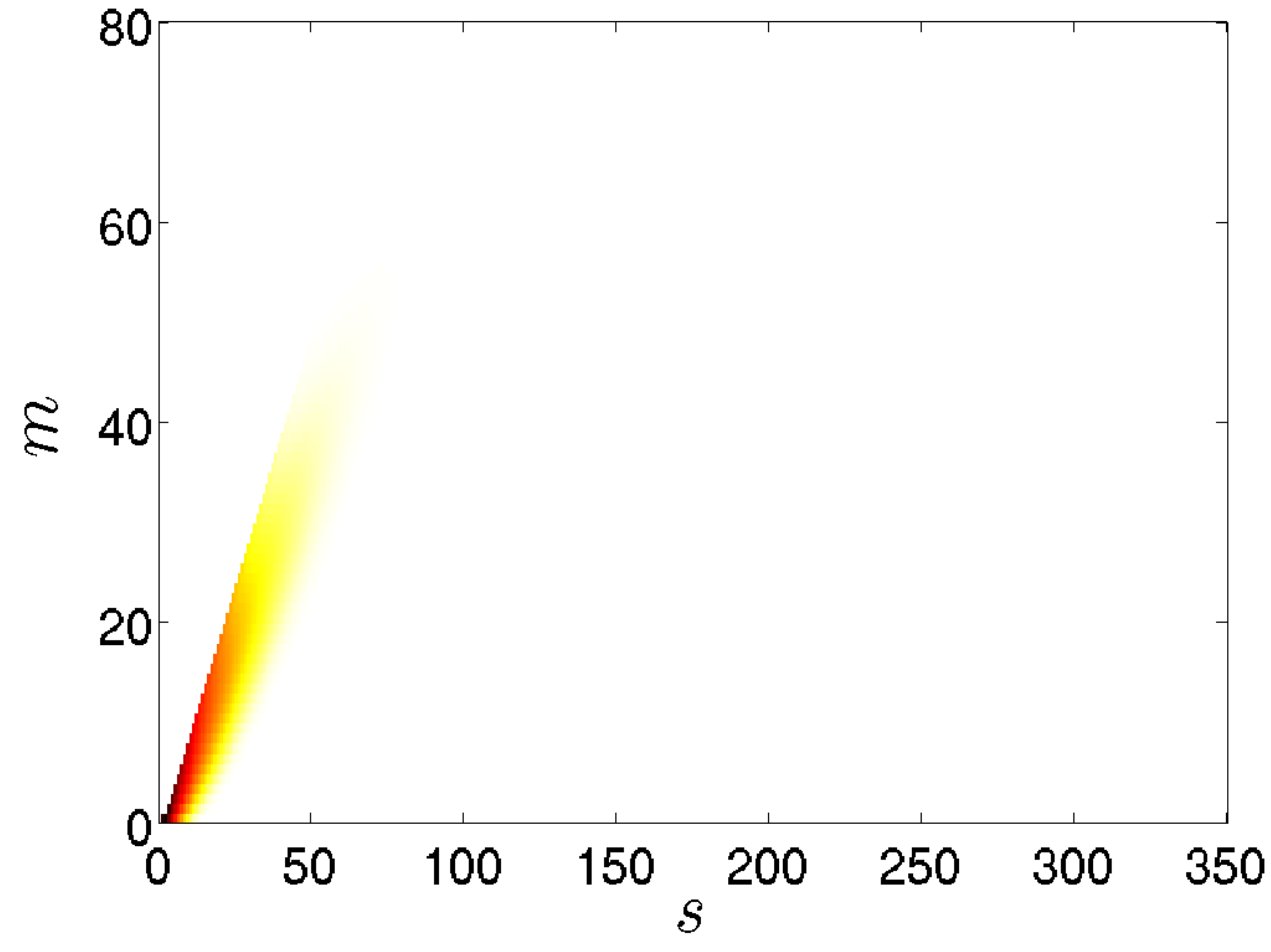} \label{subfig:plinfg2}} \hfill
\subfigure[~$g = 6$]{\includegraphics[width = .47\linewidth]{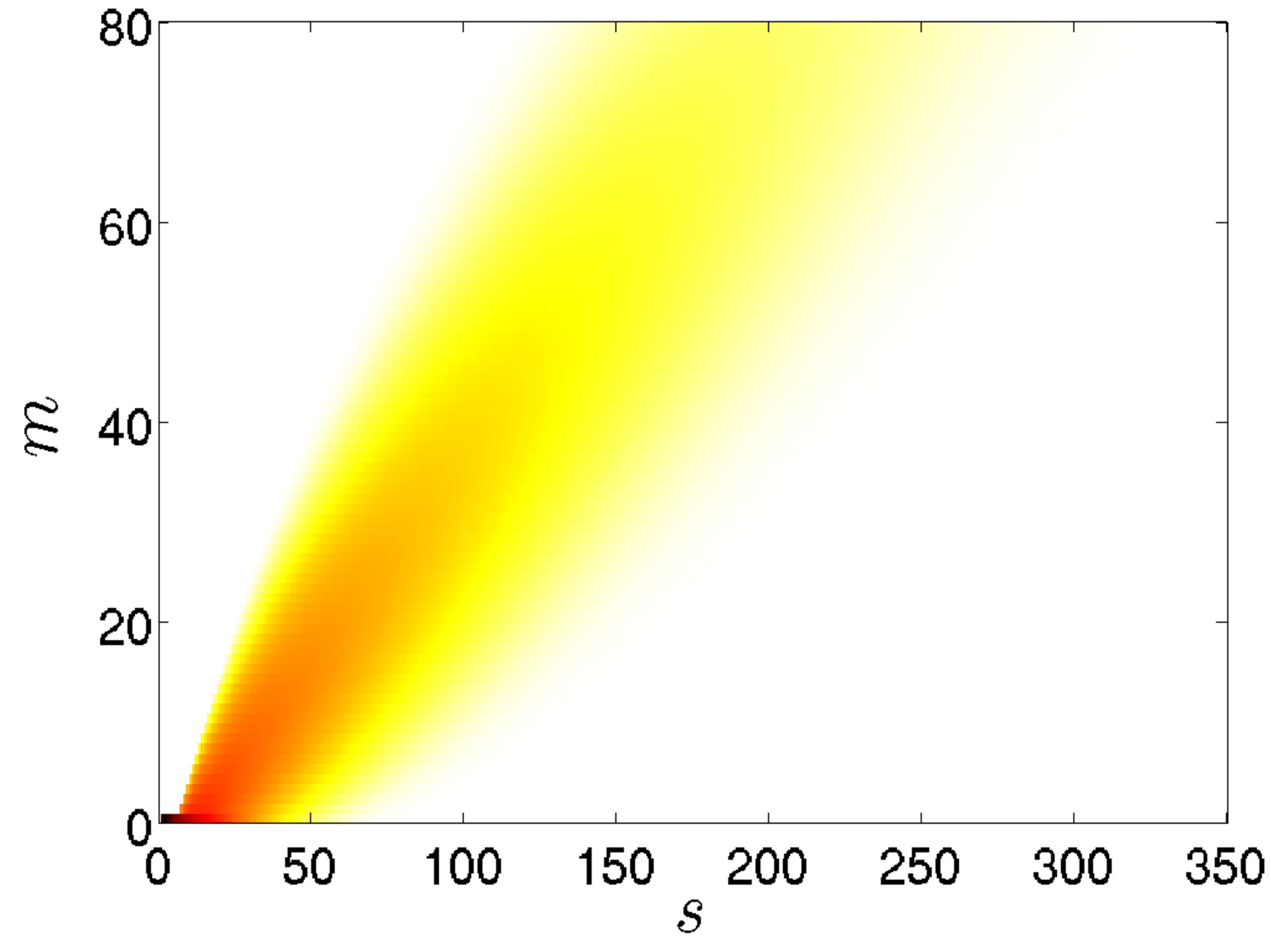} \label{subfig:plinfg6}}}
\mbox{
\subfigure[~$g = 11$]{\includegraphics[width = .47\linewidth]{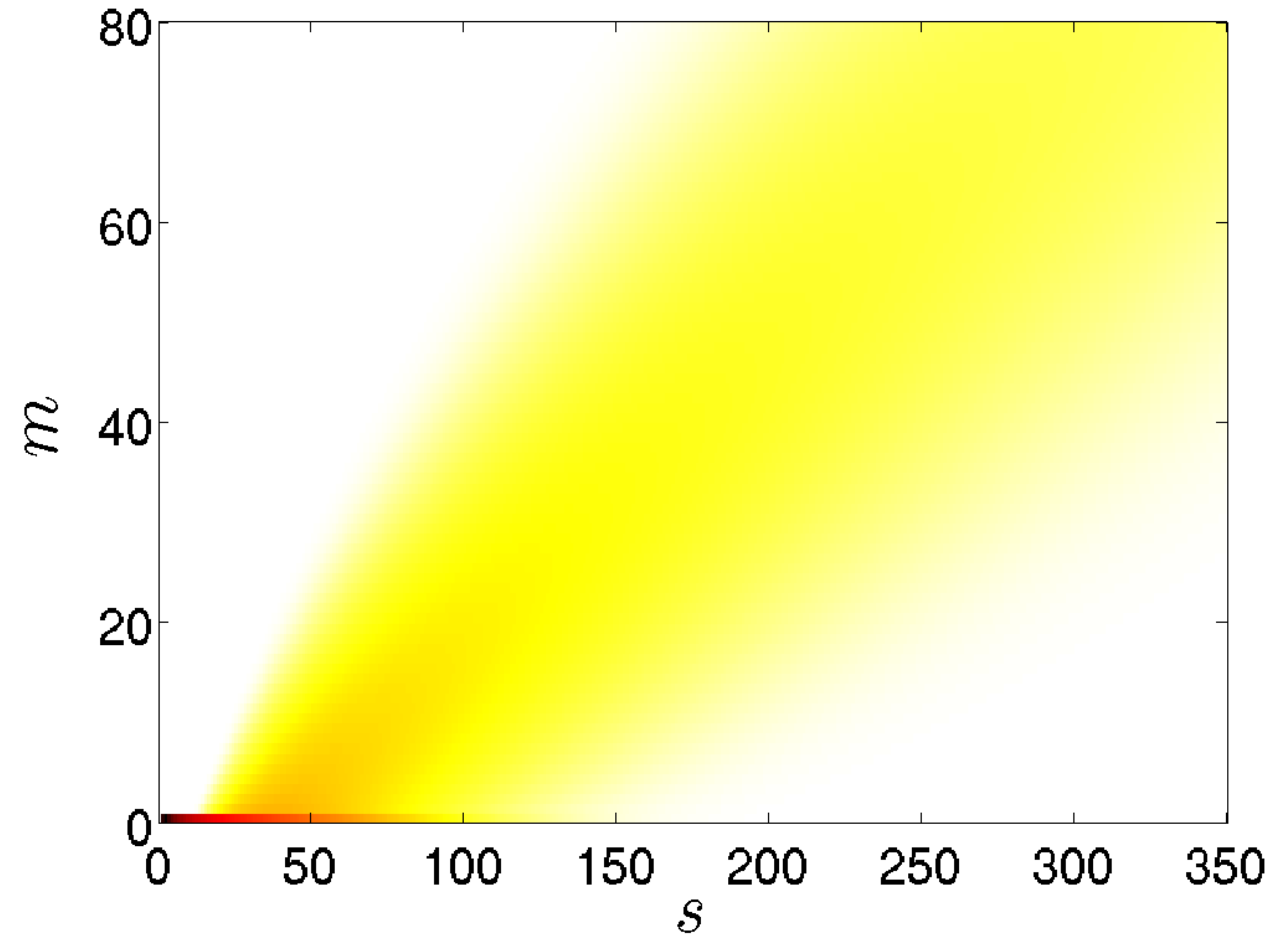} \label{subfig:plinfg11}} \hfill
\subfigure[~Final state]{\includegraphics[width = .47\linewidth]{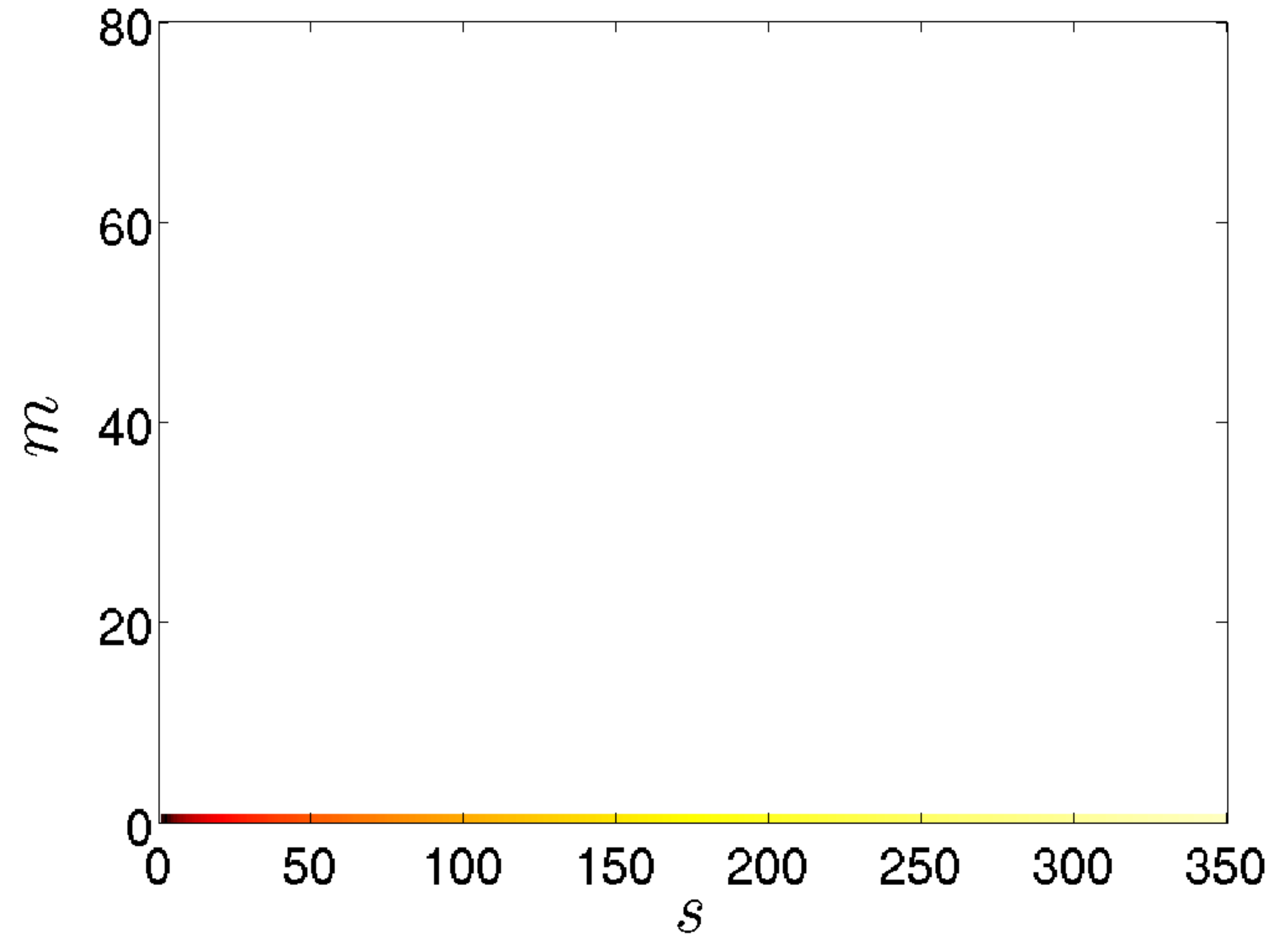} \label{subfig:plinfg100}}}
\caption{(Color online) Representation of $(s,m)$ phase space for the \emph{infinite-size} network algorithm: $s$ denotes the number of infected nodes by the end of the $g$-th generation and $m$ denotes the number of new infections that occured since the last generation. The degree distribution of the $N = 1\ 000$ nodes follows a power-law $p_k \propto k^{-\tau} {\rm e}^{-k/\kappa}$ with $\tau = 2$ and $\kappa = 5$ and the probability of transmission along an edge is $T = 0.8$ . The phase-space representations, Eq. \eqref{eq:Psi0gInf},  are displayed. for generations $2$, $6$, $11$ and the final state. \label{fig:plinfinite}}
\end{figure*}
\begin{figure}[htb]
\includegraphics[width = .95\linewidth]{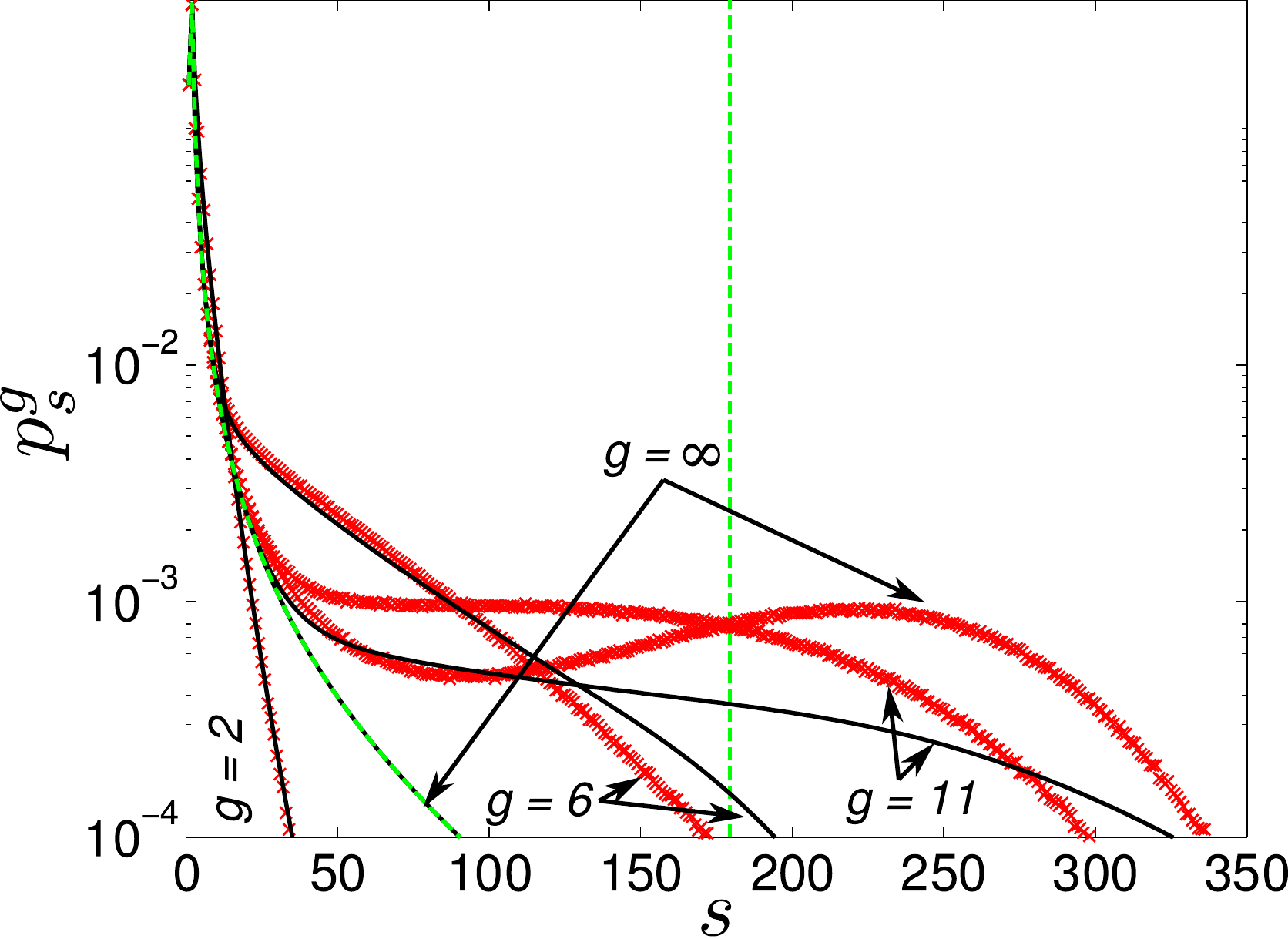}
\caption{(Color online) Projection $p_s^g$ on the $s$ axis of the $(s,m)$ phase space for the \emph{infinite-size} network algorithm: $s$ denotes the number of infected nodes by the end of the $g$-th generation and $p_s^g$ denotes the probability for $s$ to occur. The degree distribution and transmissibility of Fig. \ref{fig:plinfinite} are used and the generations shown ($2$, $6$, $11$ and final state) are also the same. The value of $p_s^g$ is plotted against $s$ (solid curves). The numerical results (dots), theoretical ``infinite-size'' outbreak distribution (dashed curves) and theoretical ``infinite-size'' epidemic size (vertical dashed lines) are also displayed. Numerical results are obtained by creating an ensemble of $10^3$ equivalent graphs, each of which is used to run $10^5$ simulations, performing $10^8$ epidemic simulations in total. The discrepancies are explained and corrected in Sec. \ref{section:discretetimefinitenetwork}. \label{fig:plinfiniteprojection}}
\end{figure}

Figure \ref{fig:plinfinite} illustrates some results of this method for a network of $N = 1\ 000$ nodes with a power law distribution $p_k \propto k^{-\tau} e^{-k/\kappa}$ with $\tau = 2$, $\kappa = 5$ and the transmissibility $T = 0.8$. The phase-space representation for generations $2$, $6$ and $11$ as well as the final state are shown. Figure \ref{fig:plinfiniteprojection} provides the corresponding projection $p_s$ on the $s$ axis (solid curves). The numerical results (crosses), theoretical ``infinite-size'' outbreak distribution (dashed curves) and theoretical ``infinite-size'' epidemic size (vertical dashed lines) are also displayed. Numerical results are obtained by creating an ensemble of $10^3$ equivalent graphs, each of which was used to run $10^5$ simulations, performing $10^8$ epidemic simulations in total \footnote{Animated versions of Figs. \protect\ref{fig:plinfinite} and \protect\ref{fig:plfinite} in form of video clips are available from the corresponding author.}.

Figure \ref{fig:plinfiniteprojection} clearly demonstrates that apart from the small-scale outbreaks, the results from the infinite-size formalism may not correctly predict the outbreak/epidemic size distribution for \emph{finite-size} networks when the fraction of the network that has been infected is no longer negligible. The remedy to this shortcoming is offered in the next section.
\section{Formalism for finite networks \label{section:discretetimefinitenetwork}}
As long as one is only interested in the \emph{initial stage} of an outbreak, the finite-size of a network has negligible effects on the dynamics of disease spread. However, the impact of finite-size effects becomes important when a sizable fraction of the network has been affected. While the size of small outbreaks is mostly governed by stochastic fluctuations, the size of the giant component (when one exists) is limited by two principal finite-size effects: the evolution over time of the degree distribution of susceptible nodes and the failure of transmission due to the impossibility of re-infection.

Since Eq. \eqref{eq:Psi0gInf} is \emph{exact} in the infinite limit, we can search for a similar form where the finite-size effects are introduced as a dependency in $s'$ and $m'$ of the degree distribution and/or of its parameters. In the following, we describe these effects and how they are introduced into the formalism described in the previous section.

It is worth noting that the finite-size effects considered here, affecting directly the dynamics on the
 network, should be distinguished from those that alter the structure of the network, for instance
through a cutoff in the degree distribution as done in \cite{pastor-satorras02b_pre}.
\subsection{Evolution of the degree distribution of susceptibles \label{subsection:evodegdist}}
As the disease progresses across the network, susceptible nodes with a higher degree of connectivity are more likely to acquire the disease than those with fewer connections. If we focus only on the degree distribution of susceptible cases, the distribution will vary over time; the portion representing high-degree susceptibles will decrease and the segment representing low-degree susceptibles will increase, to comply with normalization requirements. This variability over time has a direct effect on the ratio $z_2/z_1$ and can lower the reproduction number, $R_0$, below the threshold value of $1$. These effects can potentially cause the extinction of the disease although a high number of susceptible nodes is still remaining. This is particularly important for degree distributions in which some nodes have a degree much higher than the mean degree distribution (\emph{e.g.}, power-law distribution): the removal of these nodes will significantly impact the connectivity of the network.

\begin{figure*}[htb]
\mbox{
\subfigure[~Power-law distribution.]{\includegraphics[width = .32\linewidth]{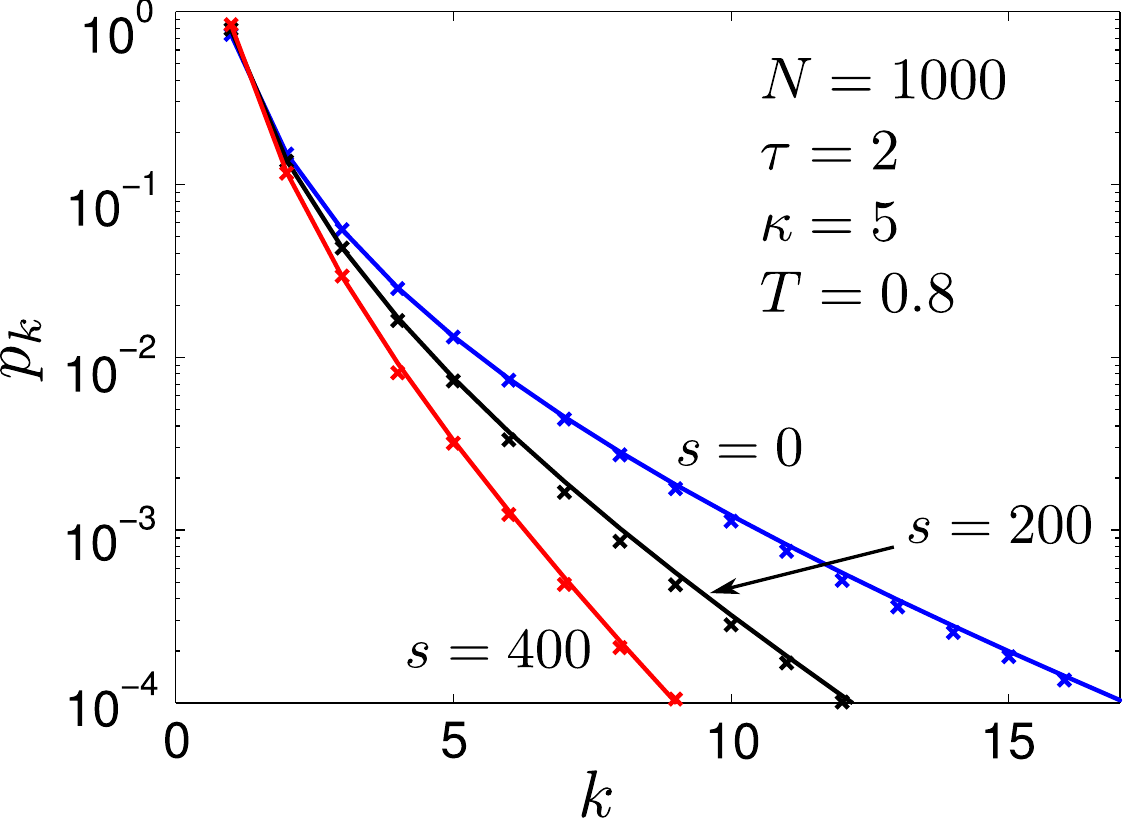} \label{subfig:EvoDegDistPl}} \hfill
\subfigure[~Binomial distribution.]{\includegraphics[width = .32\linewidth]{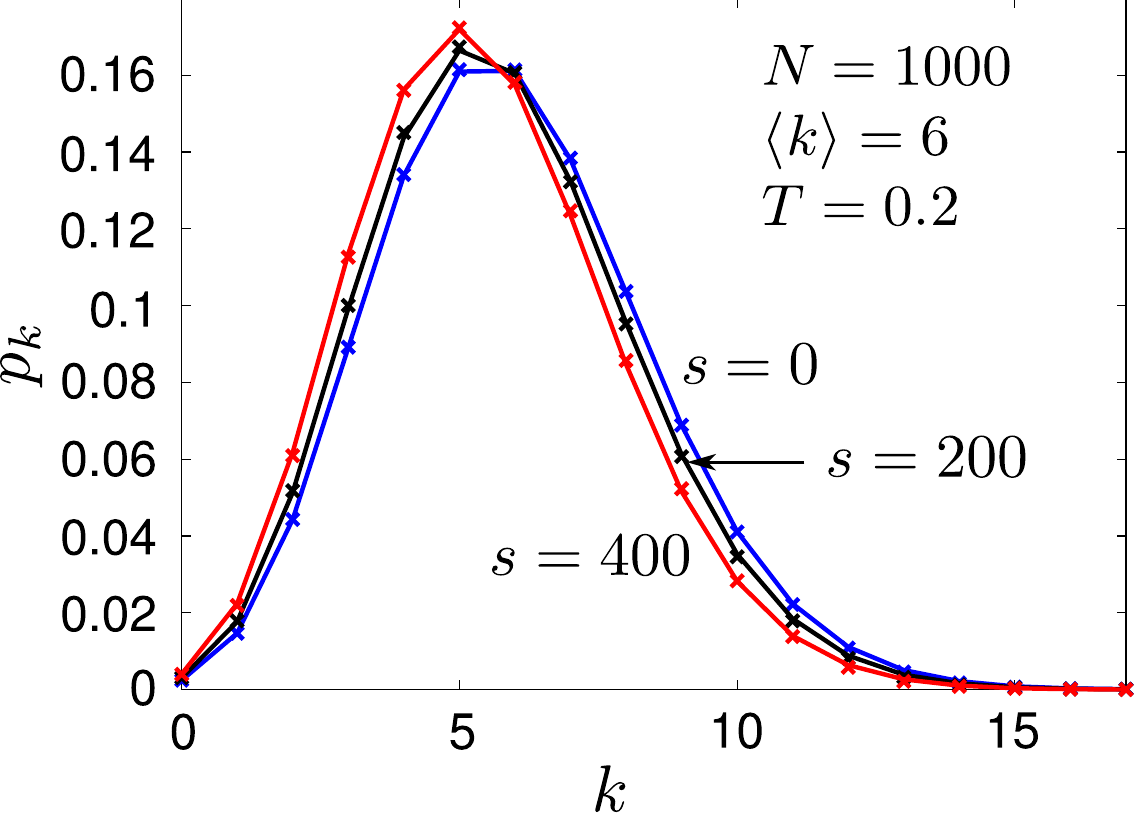} \label{subfig:EvoDegDistBino}} \hfill
\subfigure[~Bimodal distribution.]{\includegraphics[width = .32\linewidth]{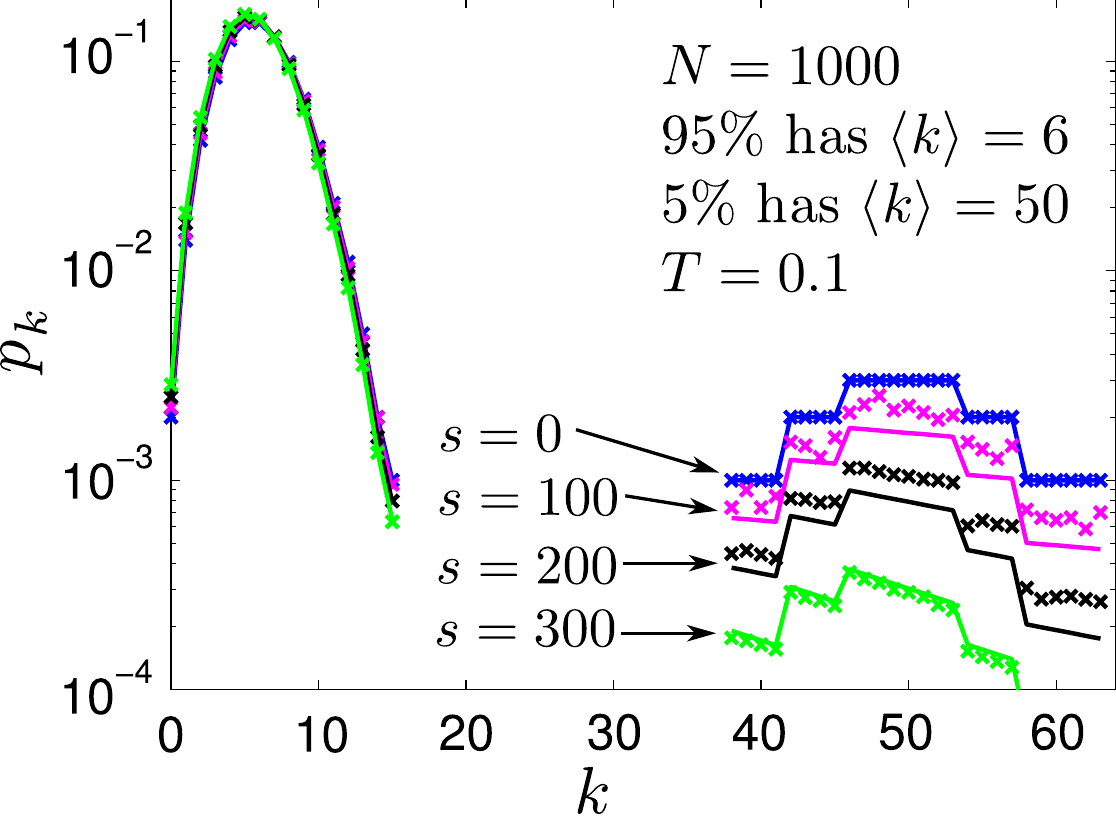} \label{subfig:EvoDegDistBimo}}}
\caption{(Color online) Time evolution of typical degree distributions representing susceptible nodes. The evolution of the degree distribution of the susceptibles is shown for 3 networks: \subref{subfig:EvoDegDistPl} the power-law distribution used in Fig. \ref{fig:plinfinite}; \subref{subfig:EvoDegDistBino} a binomial distribution with $p = 6 / N$ and $N = 1\ 000$ ; and \subref{subfig:EvoDegDistBimo} a bimodal distribution, in which the vast majority of nodes ( $95 \%$ ) has an average degree of $6$ and the rest has an average degree of $50$. The analytical (curves) and numerical (crosses,  $10^3$ equivalent graphs with $10^3$ simulations per graph) results are given for different outbreak/epidemic sizes $s$. The transmissibility values $T$ were solely used to produce numerical results. \label{fig:evodegdist}}
\end{figure*}

To take this effect into account we define the generating function for the degree distribution of the remaining susceptibles for the current size, $s$, of the outbreak/epidemic
\begin{align}
G_0^S(x;s) = \sum_k p_k^S(s) x^k \quad .
\end{align}
The mean number of susceptibles of degree $k$ is thus given by $(N - s)p_k^S(s)$. However, the actual number of susceptibles in a network characterized by $G_0^S(x;s)$ will in general be different from the mean value. Nonetheless, the difference becomes negligible in a \emph{sufficiently large} population. In this limit, each $p_k^S(s)$ can be treated as a continuous function of its parameter $s$. The assumption of a large population is less restrictive than it appears at first glance and, by comparison with the results of numerical simulations presented at the end of this section (Fig. \ref{fig:evodegdist}), we can say that it holds for reasonably small populations (\emph{e.g.} $N = 1\ 000$ for Fig. \ref{fig:evodegdist}).

Since $p_k^S(s)$ must be normalized ($\sum_k p_k^S(s) = 1$) and because the susceptibles of degree $k$ have a probability $k$ times greater of being newly infected than those of degree $1$, we can derive a differential equation system for the evolution of $p_k^S(s)$
\begin{align}
\frac{d p_k^S(s)}{ds} & = \frac{p_k^S(s)}{N-s} \left( 1 - \frac{k}{z_1^S(s)} \right) \quad , \label{eq:dpkSds}
\end{align}
where the average degree is defined by
\begin{align}
z_1^S(s) = \sum_k k\  p_k^S(s) \quad .
\end{align}

In the present dynamics, the first infection targets a random susceptible node and thus, does not affect the degree distribution. Therefore, we use the initial condition $p_k^S(1) = p_k$ (with $p_k$ being the degree distribution of the whole network) in Eq. \eqref{eq:dpkSds} to get the solution
\begin{align}
p_k^S(s) & = p_k \frac{N-1}{N-s} \left[ \theta(s) \right]^k \quad , \label{eq:pkSofs}
\end{align}
with $\theta(s)$ given by
\begin{align}
\theta(s) & = \exp\left( -\int_1^s \frac{ds'}{(N - s')z_1^S(s')} \right) \quad . \label{eq:thetas}
\end{align}
The normalization of $p_k^S(s)$ leads to the convenient expression
\begin{align}
\sum_k p_k \left[ \theta(s) \right]^k & = G_0\left(\theta(s)\right) = \frac{N-s}{N-1} \quad . \label{eq:G0thetas}
\end{align}
Moreover, Eq. \eqref{eq:pkSofs} allows us to express $G_0^S(x;s)$ in terms of the original $G_0(x)$ as
\begin{align}
G_0^S(x;s) & = \frac{N-1}{N-s} G_0\left( x \theta(s) \right) \label{eq:G0Sxs} \quad .
\end{align}

For example, using the Poisson distribution
\begin{align}
p_k & = e^{-z}z^k/k!
\end{align}
in Eq.\eqref{eq:G0thetas} gives
\begin{align}
\frac{N-s}{N-1} & = e^{-z} \sum_{k = 0}^\infty \frac{\left[ z\theta(s) \right]^k}{k!} = e^{-z} e^{z\theta(s)} \quad ,
\end{align}
from which we can easily isolate
\begin{align}
\theta(s) & = \frac{1}{z} \left[ z + \ln\left( \frac{N-s}{N-1} \right) \right] \quad .
\end{align}
It follows that the degree distribution of the \emph{susceptibles} is also a Poisson distribution 
\begin{align}
p_k^S(s) & = e^{-z_1^S(s)}\left[ z_1^S(s)\right]^k/k!
\end{align}
with an average degree given by
\begin{align}
z_1^S(s) & = z + \ln\left( \frac{N-s}{N-1} \right) \quad .
\end{align}
Inspection of this last expression reveals that $z_1^S(s)$ becomes negative when $N-s < (N-1)e^{-z}$. This limitation is due to the fact that the large population assumption is no longer respected: it implies the presence of $(N-1)e^{-z}$ nodes of degree zero, but these nodes cannot be infected by the process leading to Eq. \eqref{eq:dpkSds}. Nevertheless, the probability of an epidemic reaching such high values of $s$ typically vanishes in most realistic cases. Table \ref{table:thetas} compiles the analytic forms of $\theta(s)$ for some typical distributions. When a closed form satisfying Eq. \eqref{eq:G0thetas} cannot be found, the quantity $p_k^S(s)$ can nevertheless be derived numerically for each pair of $k$ and $s$.

\begin{table*}[htb]
\caption{Expression for $\theta(s)$ for some commonly used degree distributions. \label{table:thetas}}
\begin{ruledtabular}
\begin{tabular}{lll}
           & Degree distribution                  & Expression for $\theta(s)$ \\
\hline
& & \\
Poisson  & $\displaystyle{p_k = \frac{e^{-z}}{k!} z^k}$            &
$\displaystyle{\theta(s) = \frac{1}{z}\left[ z + \ln\left( \frac{N-s}{N-1} \right) \right]}$      \\
& & \\
Binomial & $\displaystyle{p_k = \binom{N}{k} p^k (1-p)^{N-k}}$ &
$\displaystyle{\theta(s) = \frac{1}{p}\left[ \left(\frac{N-s}{N-1}\right)^{1/N} + p - 1 \right]}$ \\
& & \\
Exponential& $\displaystyle{p_k = (1 - e^{-1/\kappa})e^{-k/\kappa}}$ &
$\displaystyle{\theta(s) = \frac{N-1 - (s-1)e^{e^{1/\kappa}}}{N-s}}$ \\
& & \\
Power law & $\displaystyle{p_k = \frac{k^{-\tau}e^{-k/\kappa}}{\Li_\tau\left( e^{-1/\kappa} \right)}}$ for $k \ge 1$ &
$\displaystyle{\Li_\tau\left( e^{-1/\kappa} \theta(s) \right) = \frac{N-s}{N-1} \Li_\tau\left( e^{-1/\kappa} \right)}$ \\
& &
\end{tabular}
\end{ruledtabular}
\end{table*}

Figure \ref{fig:evodegdist} shows the variation of the degree distributions of susceptible nodes for 3 networks: \subref{subfig:EvoDegDistPl} the power-law distribution used in Fig. \ref{fig:plinfinite}; \subref{subfig:EvoDegDistBino} a binomial distribution with $p = 6/N$ and $N = 1\ 000$; and \subref{subfig:EvoDegDistBimo} a bimodal distribution, in which the vast majority of nodes ($95 \%$) has an average degree of $6$ and the rest has an average degree of $50$. This latter network is particularly interesting as it may correspond to realistic settings such as hospitals, schools or shopping malls. Although slight deviations are caused by the underlying assumptions, there is very good agreement between numerical and analytical results.

Once $G_0^S(x;s)$ is known, we can show that the degree distribution of the susceptibles in the previous generation is given by $G_0^S(x;s-m)$. Again using mean-value considerations, we obtain the degree distribution, $G_0^I(x;s,m)$, of those that became infectious in the last generation, \emph{i.e.}
\begin{multline}
m G_0^I(x;s,m) = \bigl( N - (s-m) \bigr) G_0^S(x;s-m) \\ - \bigl( N - s \bigr) G_0^S(x;s) \quad ,
\end{multline}
as
\begin{multline}
G_0^I(x;s,m) = \\ (N - 1)\left[ \frac{G_0\left( x \theta(s-m) \right) - G_0\left( x \theta(s) \right)}{m} \right] \quad .
\end{multline}
The excess degree of the currently infectious nodes is therefore generated by
\begin{align}
\tilde{G}_g(x;s,m) & = \begin{cases}
                       G_0(x)                 & \text{($g = 0$)}  \\
                       \displaystyle{\frac{G_0^I(x;s,m)}{x}} & \text{($g \ge 1$)}
                       \end{cases} \quad . \label{eq:Gtildeg}
\end{align}
This distribution can then be used in Eq. \eqref{eq:Psi0gInf} as a substitute for $G_g(x)$ when finite-size effects cannot be neglected.

\subsection{Additional loss of transmissions \label{subsection:AddLossTrans}}
For networks of finite size, it is no longer possible to completely neglect the effect of closed loops on the dynamics of outbreaks. Indeed, it is possible that some of the neighbors of a newly infected node have previously been infected and, for dynamics where re-infection is impossible, this implies fewer new infections than would have been predicted in an infinite network. Similarly, links between two infectious nodes or links from more than one infectious node to the same susceptible node also reduce the number of new infections.

Furthermore, since the pair $s'$ and $m'$ completely characterizes the ``state'' of the system in an infinite network, it still carries a lot of information about the corresponding state in finite-size networks. We thus make the assumption that $s'$ and $m'$ are a sufficient basis to incorporate the finite-size effects, \emph{i.e.} the loss of transmissions in the finite network that would have occurred in an infinite one.

Our main step is to seek a ``mean field'' approximation (where every parameter other than $s'$ and $m'$ is assumed to take its mean value) to the ratio
\begin{align}
\rho_{s'm'} = \left\langle \frac{\langle \tilde{m} \rangle}{m} \right\rangle_{s'm'} \label{eq:rhoratiodef}
\end{align}
of mean number of transmissions $\tilde{m}$ that \emph{actually happen} in the \emph{finite} network to the number of transmissions $m$ that \emph{would have occurred} in an \emph{infinite} one. If each of the $m$ infinite network transmissions is treated as having the \emph{independent} probability $1 - \rho_{s'm'}$ of being lost (a probability $\rho_{s'm'}$ of occurring) in the finite network, there is then a probability
\begin{align}
\binom{m}{\tilde{m}} (\rho_{s'm'})^{\tilde{m}} (1 - \rho_{s'm'})^{m - \tilde{m}} \label{eq:distmtilde}
\end{align}
that $\tilde{m}$ transmissions occur in the finite network. Similarly, a subgroup $l$ of the $m$ infinite network transmissions will
contribute $\tilde{l}$ to the $\tilde{m}$ finite network transmissions with probability
\begin{align}
\binom{l}{\tilde{l}} (\rho_{s'm'})^{\tilde{l}} (1 - \rho_{s'm'})^{l - \tilde{l}} \label{eq:distltilde} \quad .
\end{align}

Although these transmissions are not exactly independent events, the independence assumption is a good approximation when the system has many degrees of freedom, \emph{i.e.}, it holds when $m$ is small compared to the number of susceptibles of nonzero degree. The probability distributions for $\tilde{m}$ (Eq. \eqref{eq:distmtilde}) and $\tilde{l}$ (Eq. \eqref{eq:distltilde}) therefore hold unless most nodes of the network have been infected, quite unlikely for example in realistic epidemiological applications. One of the first effects of a non-negligible correlation would typically be a reduction of the variance of the distributions of $\tilde{m}$ and $\tilde{l}$.

Appendix \ref{section:rhosm} provides an expression for $\rho_{s'm'}$ under assumptions of continuity (a differential equation approach similar to the previous section) for a sizable population, large enough to obtain meaningful mean values. All but one of these mean values are relatively easy to calculate, the remaining one is derived in Appendix \ref{section:nRsm} to complete the task.
\subsection{Phase-space representation in finite-size networks}
We now combine the degree distribution of infectious nodes
\begin{align}
\tilde{G}_{g-1}(x;s',m') & = \sum_k \tilde{p}_k(s',m') x^k
\end{align}
obtained in Sec. \ref{subsection:evodegdist} with the distribution of $\tilde{l}$ to obtain the finite-size network counterpart of Eq. \eqref{eq:G01pxm1T}
\begin{widetext}
\begin{align}
\sum_{\tilde{l}=0}^\infty \sum_{l=\tilde{l}}^\infty \sum_{k=l}^\infty \tilde{p}_k(s',m') \binom{k}{l} T^l (1-T)^{k - l} \binom{l}{\tilde{l}} (\rho_{s'm'})^{\tilde{l}} (1 - \rho_{s'm'})^{l - \tilde{l}} x^{\tilde{l}} & = \tilde{G}_{g-1}\left(1 + (x-1) T \rho_{s'm'} ; s',m' \right) \quad .
\end{align}
\end{widetext}
This is now the generating function for the number of new infections caused by a single infectious node in a finite network when the state of the disease is characterized by $s'$ and $m'$ (at generation $g$). Defining an effective transmissibility $\tilde{T}_{s'm'} = T \rho_{s'm'}$, the forward recurrence relation, Eq. \eqref{eq:Psi0gInf}, is then replaced by a new expression of the same structure
\begin{multline}
\widetilde{\Psi}_0^g(x,y) =  \sum_{s',m'} \widetilde{\psi}_{s'm'}^{g-1}\  x^{s'} \\ \times \left[ \tilde{G}_{g-1}\left( 1 + (xy - 1)\tilde{T}_{s'm'} ; s',m' \right) \right]^{m'} \quad , \label{eq:Psi0gFin}
\end{multline}
where the evolution of the degree distribution is taken into account by the change $G_{g-1}(x) \rightarrow \tilde{G}_{g-1}(x;s',m')$ while the additional losses of transmissions are introduced by the replacement $T \rightarrow \tilde{T}_{s'm'}$.

\begin{figure*}[htb]
\mbox{
\subfigure[~$g = 2$]{\includegraphics[width = .48\linewidth]{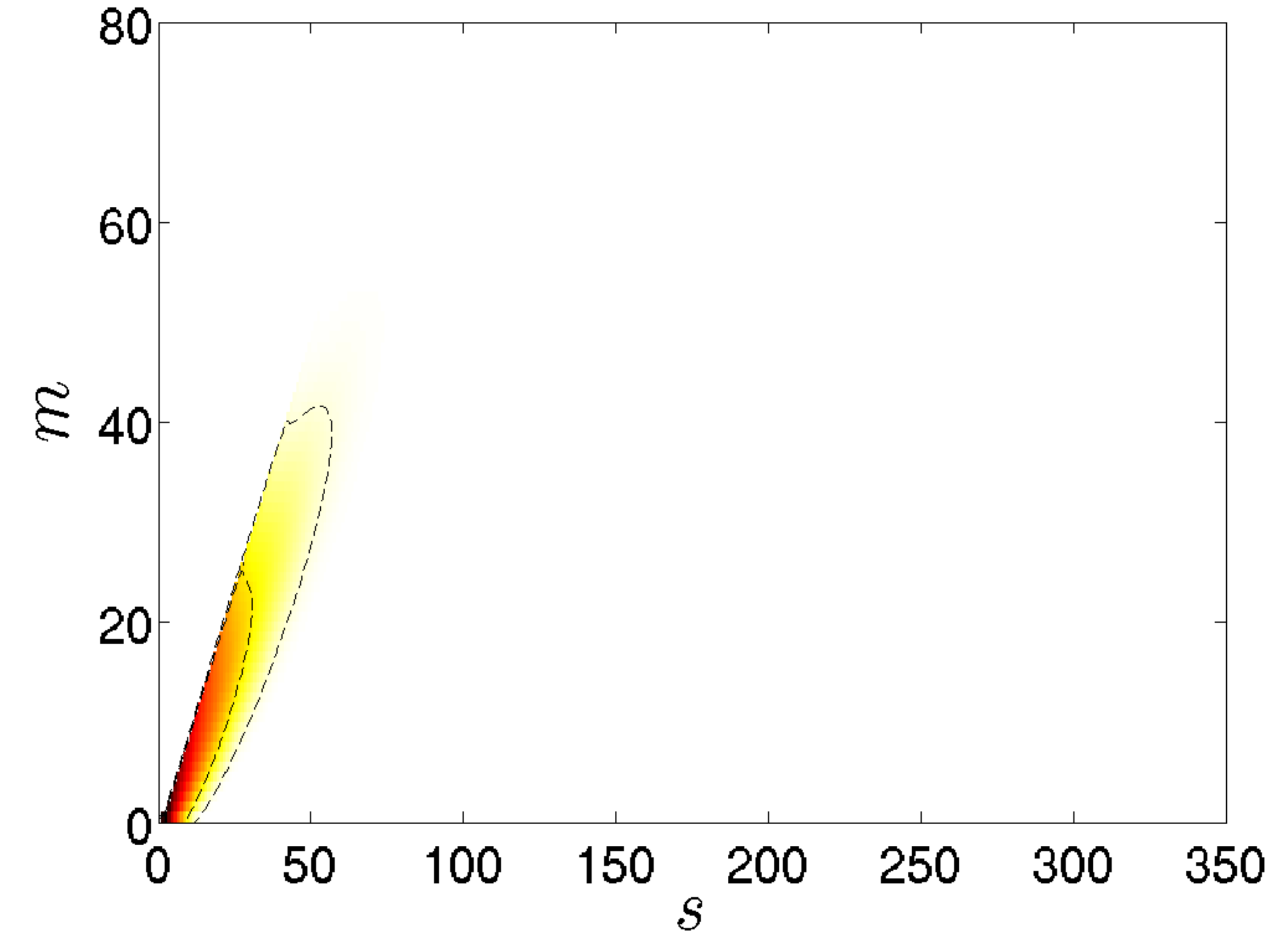} \label{subfig:plfing2}} \hfill
\subfigure[~$g = 6$]{\includegraphics[width = .48\linewidth]{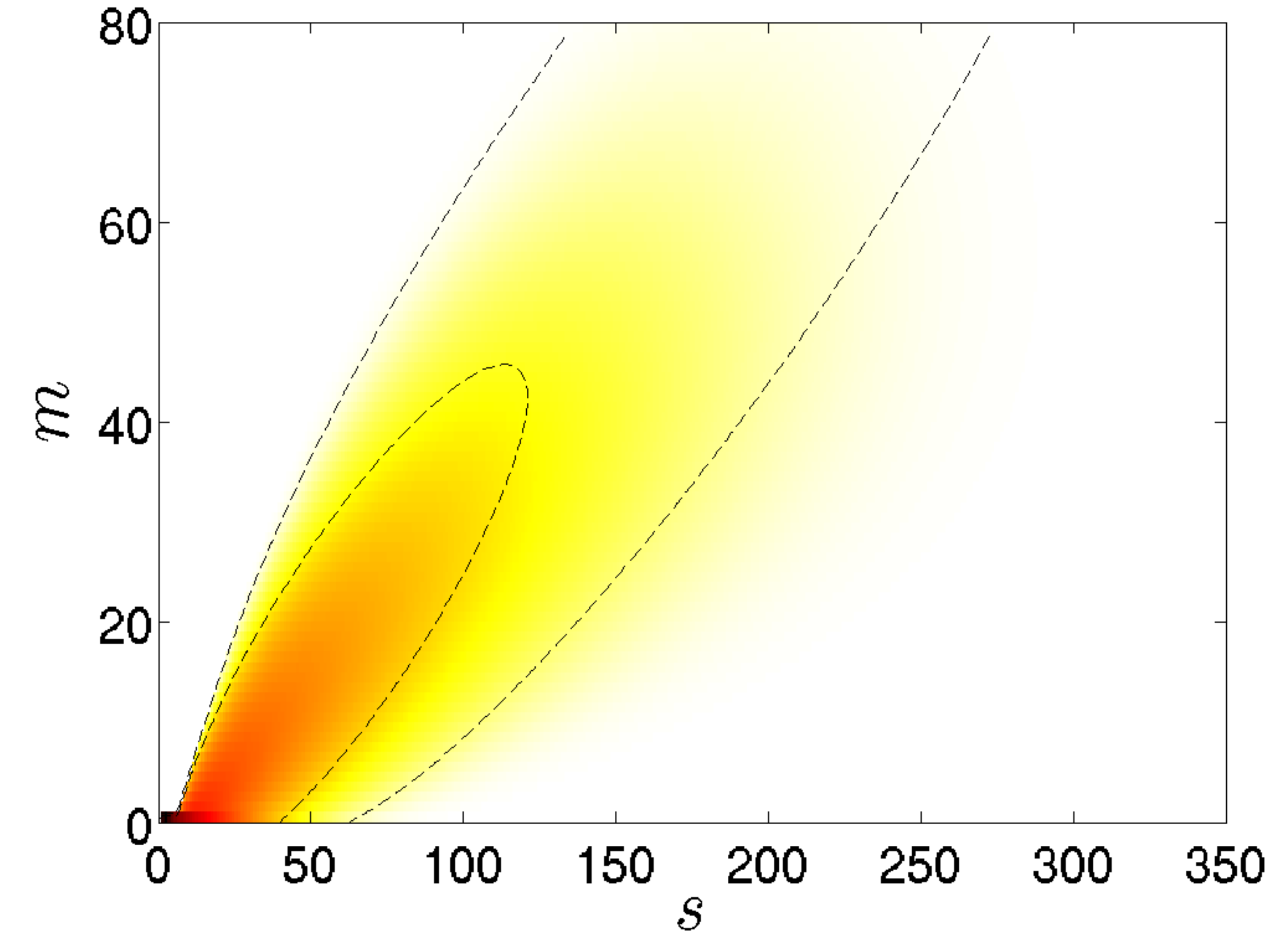} \label{subfig:plfing6}}}
\mbox{
\subfigure[~$g = 11$]{\includegraphics[width = .48\linewidth]{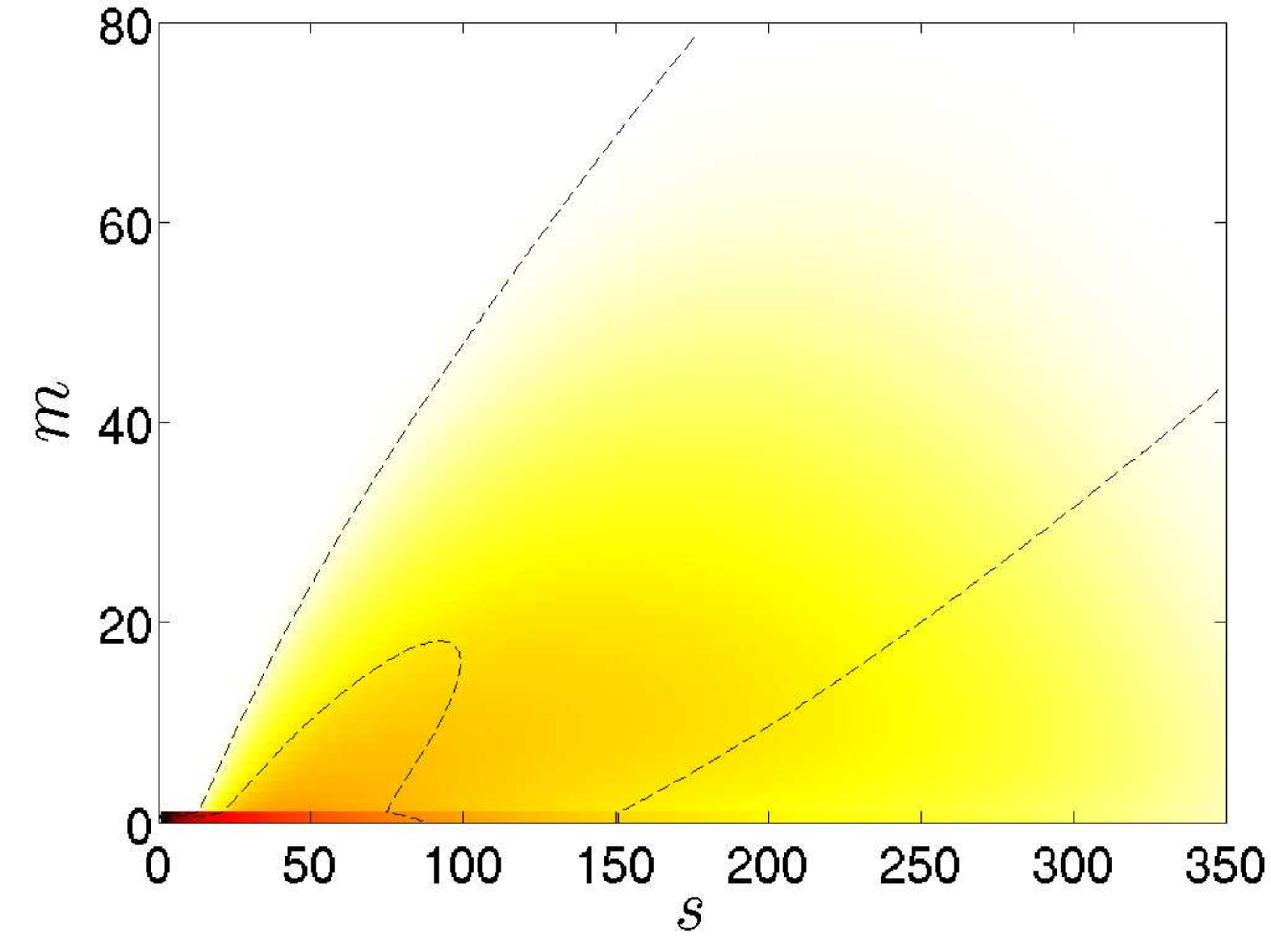} \label{subfig:plfing11}} \hfill
\subfigure[~Final state]{\includegraphics[width = .48\linewidth]{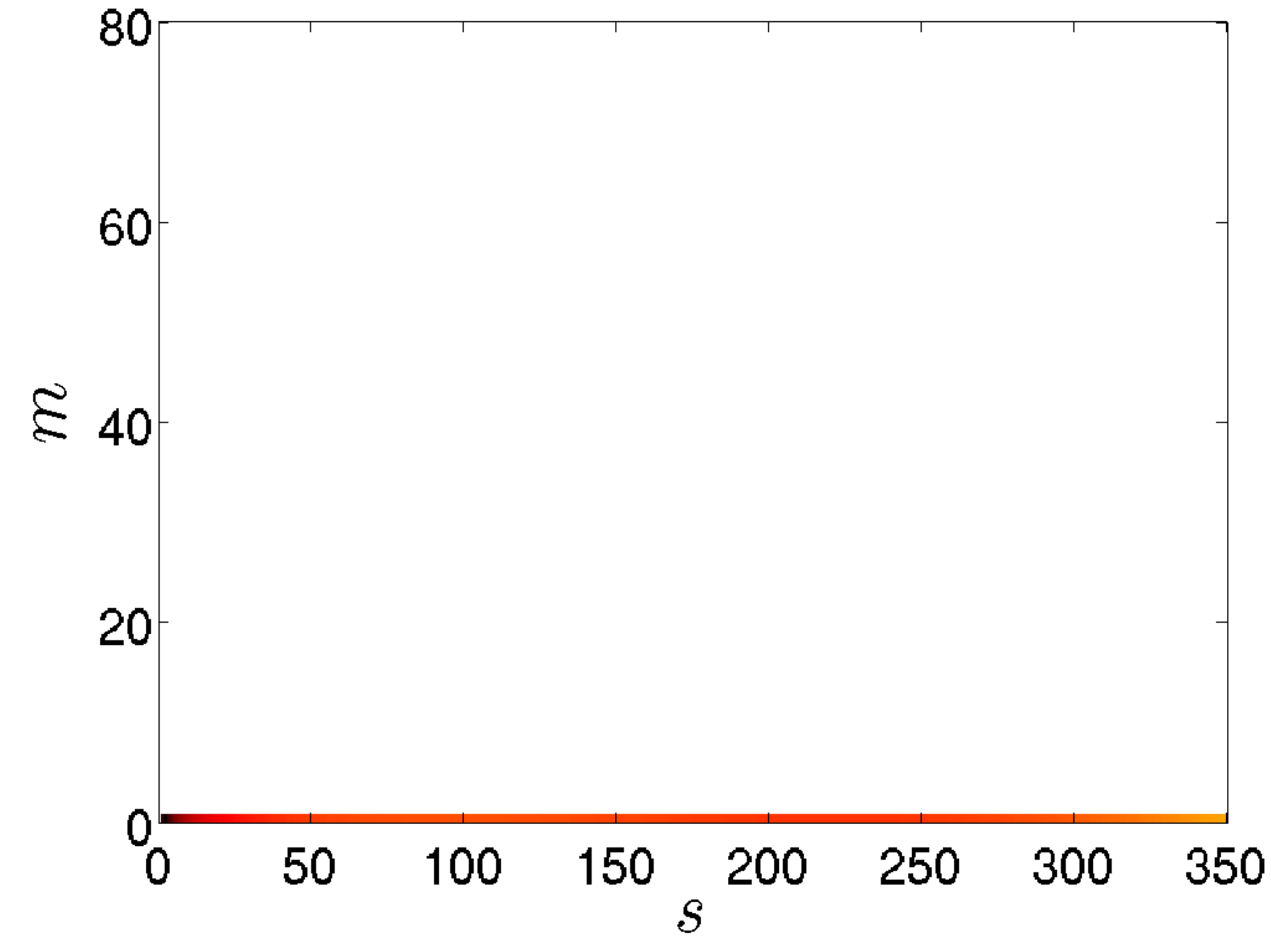} \label{subfig:plfing100}}}
\caption{(Color online) Representation of $(s,m)$ phase space for the \emph{finite-size} network algorithm: $s$ denotes the number of infected nodes by the end of the $g$-th generation and $m$ denotes the number of new infections that occurred since the last generation. The degree distribution, transmissibility and symbols are the same as for Fig. \ref{fig:plinfinite} except that the finite-size algorithm \eqref{eq:Psi0gFin} is used for the calculations. For comparison, the dashed curves are contour plots of the previous results for infinite-size network (Fig. \ref{fig:plinfinite}, Eq. \eqref{eq:Psi0gInf}). \label{fig:plfinite}}
\end{figure*}
\begin{figure}[htb]
\includegraphics[width = .95\linewidth]{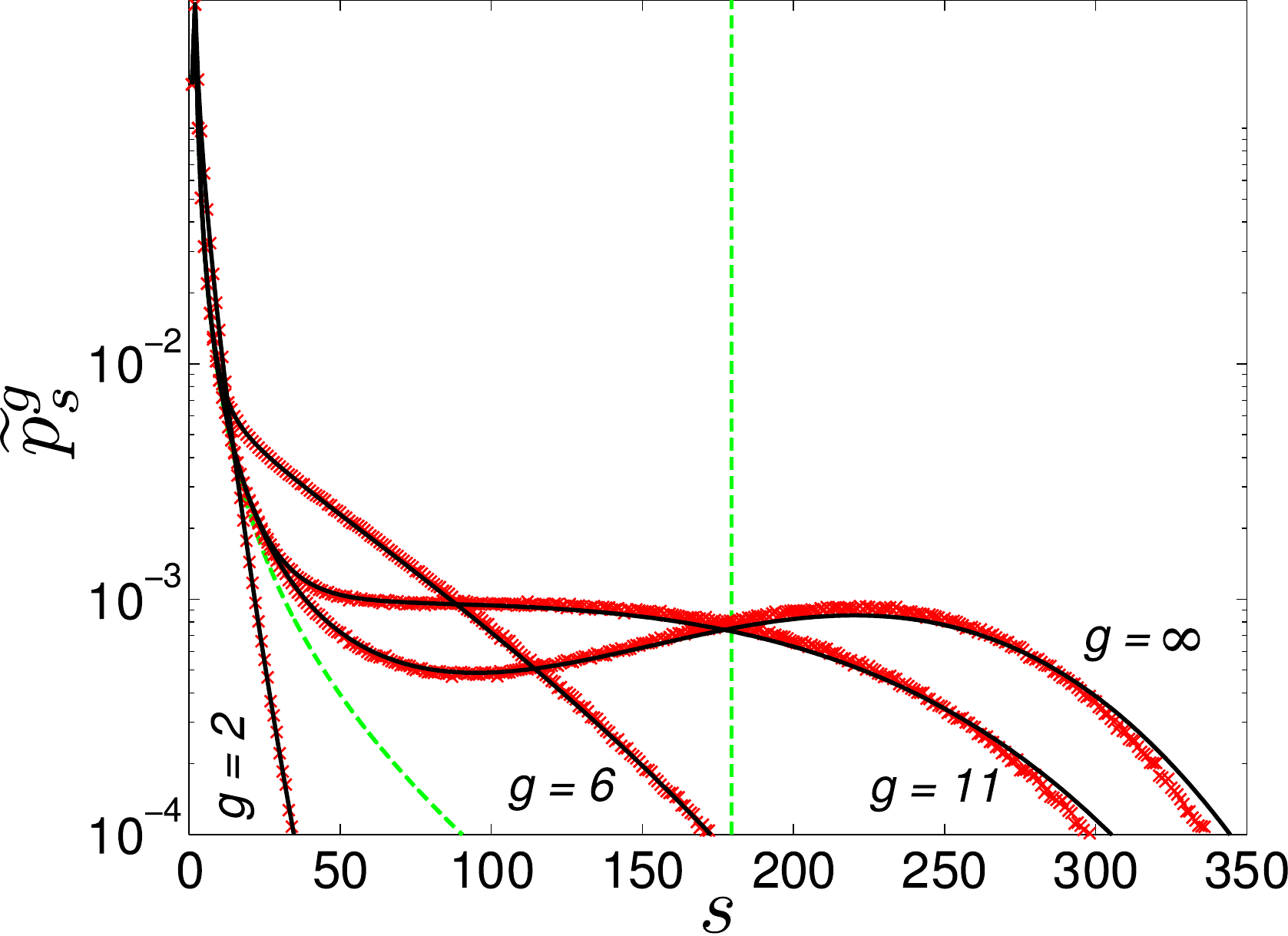}
\caption{(Color online) Projection $\widetilde{p}_s^g$ on the $s$ axis of the $(s,m)$ phase space for the \emph{finite-size} network algorithm: $s$ denotes the number of infected nodes by the end of the $g$-th generation and $\widetilde{p}_s^g$ denotes the probability for $s$ to occur. The degree distribution and transmissibility of Fig. \ref{fig:plinfinite} are used and the symbols used are those of Fig. \ref{fig:plinfiniteprojection}. The results produced by the finite-size algorithm are in very good agreement with the numerical simulations ($10^3$ equivalent graphs with $10^5$ simulations per graph) over the entire range of possible outbreak/epidemic sizes. \label{fig:plfiniteprojection}}
\end{figure}
Figure \ref{fig:plfinite} depicts the phase-space representation for the finite-size power-law network, using the finite-size algorithm of Eq. \eqref{eq:Psi0gFin} instead of the infinite size approach of Eq. \eqref{eq:Psi0gInf}. The system studied is identical with that of Fig. \ref{fig:plinfinite} and for comparison the infinite-size calculations are superimposed on the new results. Figure \ref{fig:plfiniteprojection} presents the corresponding projections on the $s$ axis. There is a major improvement in the agreement with the numerical simulations for all generations and the final state. Despite the approximations, and the small size of the network ($N = 1\ 000$), this agreement makes us confident that we have captured the major part of the finite-size effects.

Furthermore, while the infinite-size formalism produces a single number, $S$, for the giant component size (represented by the dashed vertical line in Fig. \ref{fig:plfiniteprojection}), the finite-size formalism produces the \emph{whole} probability distribution of sizes above the epidemic threshold. More on this in the next section.
\section{Futher Results and Discussion \label{section:discussion}}
In what follows, we examine in succession the results of the finite versus infinite formalism
on and around the percolation threshold, make a comparison with recent dynamical models of disease
propagation, establish the relationship with the reproductive number of epidemiology and briefly
discuss possible avenues of improvements concerning clustering and continuous time evolution.
\subsection{Behaviour around percolation threshold \label{subsection:behathreshold}}
\begin{figure}[htb]
\includegraphics[width = .95\linewidth]{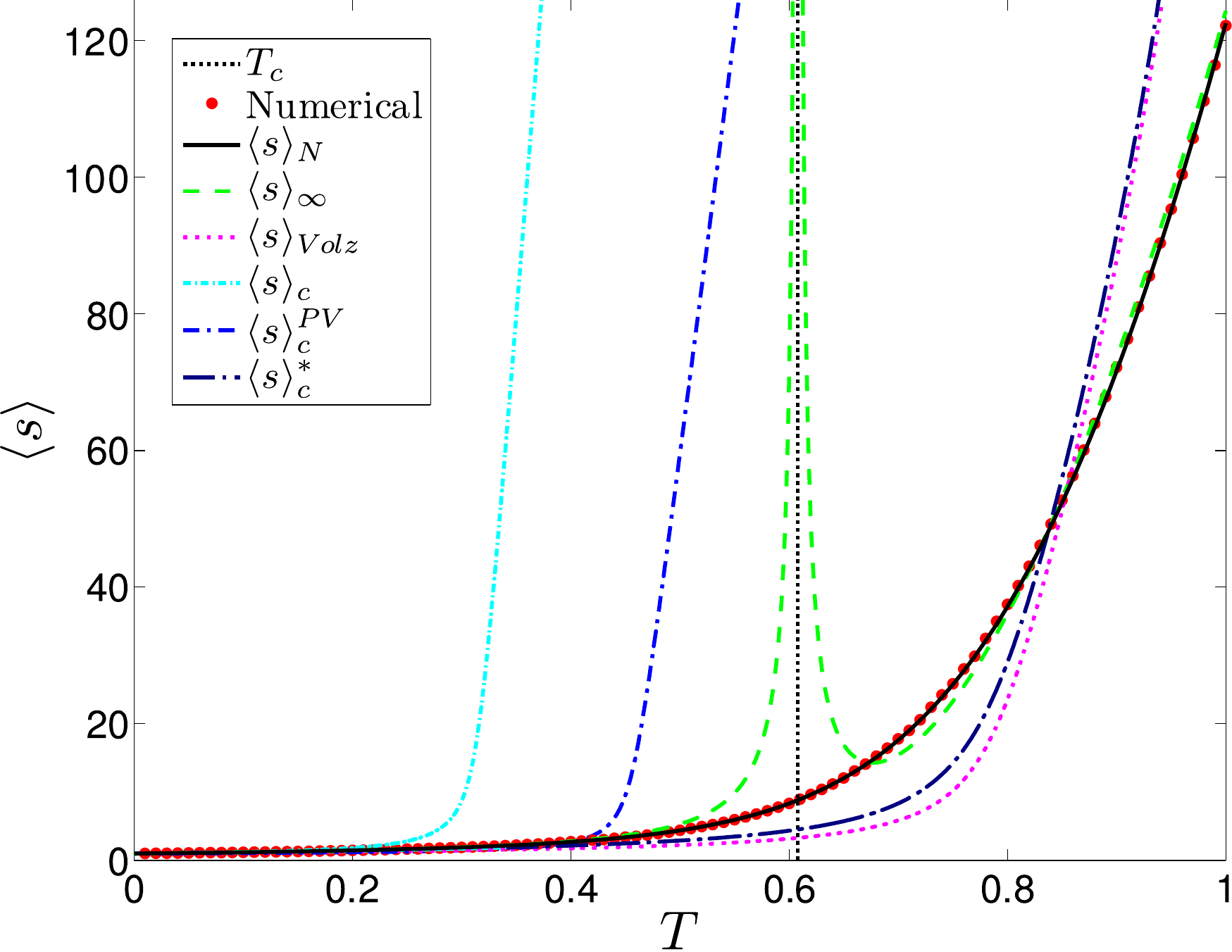}
\caption{(Color online) Average size of outbreaks and/or epidemics for the power-law network of Fig. \ref{fig:plinfinite} ($N = 1\ 000$) as a function of transmissibility, $T$. The infinite-size formalism provides the position of critical transmissibility, $T_c \simeq 0.6080$, above which epidemics can occur (vertical dotted line), as well as the expected size of outbreaks or epidemics $\langle s \rangle_\infty$ (Eq. \eqref{eq:meansInf}, dashed curve). In contrast to the divergence at $T_c$ in the infinite-size formalism, numerical simulations (circles, $10^3$ equivalent graphs with $10^3$ simulations per graph
for 100 different values of $T$) 
show a smooth monotonic increase of $\langle s \rangle$. The prediction of the new finite-size algorithm 
($\langle s \rangle_N$, Eq. \eqref{eq:meansFin}, solid line) is in perfect agreement with numerical results. Four other curves provide the results of the models presented in Sec \ref{subsection:varietyoutcome}. \label{fig:plbehaviourthreshold}}
\end{figure}
We recall first that the current paper has been strongly influenced by the previous works of \cite{newman01_pre,newman02_pre}. Their formalism allows for the calculation of the whole distribution of outbreak/epidemic size and probability at the final state of the dynamics ($g \rightarrow \infty$ in the language of this paper) for infinite networks. It is based on the \emph{pgf}s $H_0(x)$ and $H_1(x)$ related to $G_0(x)$ and $G_1(x)$ of Eqs. \eqref{eq:G0} and \eqref{eq:G1} by the relations
\begin{align}
H_1(x) & = x G_1\left( 1 - \bigl( 1 - H_1(x) \bigr) T \right) \label{eq:NewmanH1} \\
H_0(x) & = x G_0\left( 1 - \bigl( 1 - H_1(x) \bigr) T \right) \label{eq:NewmanH0} \quad .
\end{align}
In particular, the \emph{pgf} $H_0(x)$ generates the distribution of size for outbreaks from which the average size of outbreak/epidemic for infinite networks can be expressed as 
\begin{equation}
\langle s \rangle_\infty = H_0(1) H_0'(1) + N \, \bigl( 1 - H_0(1) \bigr)^2 \quad . \label{eq:meansInf}
\end{equation}
Note also that above threshold the fraction of nodes belonging to the giant component is given
by $S= 1 -H_0(1)$; below threshold $H_0(1)= 1$ as it should.

Since Sec. \ref{section:InfiniteNetFormalism} is a direct extension of this formalism to discrete time evolution, the values it provides all converge to that of \cite{newman02_pre} at the final state of the dynamics. Formally, this is no surprise since as $g \to \infty$, the recurrence relationship \eqref{eq:Psi0gInf} approaches the self-consistency expressions \eqref{eq:NewmanH1} and \eqref{eq:NewmanH0}. In this asymptotic limit, $H_0(x) = \Psi_0^\infty(x,1)$. Similar remarks hold for the work of Marder \cite{marder07_pre}, another extension of \cite{newman02_pre} to discrete time evolution.

By contrast, the formalism of Sec. \ref{section:discretetimefinitenetwork} takes into account finite-size effects. Instead of Eq. \eqref{eq:meansInf}, the average size of an outbreak/epidemic is now provided by
\begin{equation}
\langle s \rangle_N = \sum_{s,m} s \, \widetilde{\psi}_{sm}^\infty \label{eq:meansFin}
\end{equation}
since the distribution $\widetilde{\psi}_{sm}^\infty$ is always properly normalized with 
$\widetilde{p}_s^\infty= \sum_m \widetilde{\psi}^\infty_{sm}$ and $\sum_s \widetilde{p}_s^\infty= 1$.

An important feature of the new formalism is its ability to capture the finite size of an outbreak in the vicinity of the critical transmissibility, $T_c$, which separates small outbreak and large-scale epidemic zones. This notion is depicted in Fig. \ref{fig:plbehaviourthreshold} where we consider the same power-law network for different transmissibilities $T$ ($T_c \simeq 0.6$). In an infinite-size network, we observe a ``divergence'' in the outbreak/epidemic size, $\langle s \rangle_\infty$, just around the transmissibility threshold. However, numerical simulations on finite-size networks (and similarly real-life outbreaks) never exhibit such divergence. Figure \ref{fig:plbehaviourthreshold} demonstrates that the finite-size formalism can accurately retrieve the size of outbreaks on and around the transmissibility threshold. Away from threshold, infinite and finite methods agree with the simulations, but only the finite-size approach covers smoothly the complete range of
 transmissibilities without unrealistic divergence.
\subsection{Importance of the variety of outcomes \label{subsection:varietyoutcome}}
It would be informative to compare our method with some other propagation approaches on a structured network. The chosen models all perform some ``mean field'' approximations that prevent them from taking into account the wide variety of possible outcomes, which can span the whole range from small outbreaks to large-scale epidemics. In effect, these models can only provide mean values and the quality of the resulting averages is sometimes difficult to estimate.

For the purpose of comparison, all the models are defined for the same network ensemble as the one presented in Sec. \ref{section:InfiniteNetFormalism} and the quantity that we will focus on is the average size of outbreaks/epidemics at the asymptotic final state of the dynamics.
\subsubsection{Volz (2008): A network-centric approach}
Close in spirit with the present contribution, but technically very different, is the recent approach derived by Volz \cite{volz08_jmathbio}. The formalism is based on network-centric quantities (such as edges linking susceptible nodes to infectious nodes or to other susceptible nodes) in order to introduce time evolution and finite-size effects. Although it is also based on the \emph{pgf} $G_0(x)$ (Eq. \eqref{eq:G0}), it uses a set of differential equations to track the evolution of mean values. The network-centric quantities are later converted to node-centric quantities, such as the number of infected or susceptible nodes. The asymptotic ($t \to \infty$) average size of an outbreak/epidemic we are interested in can be cast in the following form
\begin{equation}
\langle s \rangle_{\text{Volz}} = N \, \left( 1 - G_0\bigl( \theta_V( \infty ) \bigr) \right)
\end{equation}
where $\theta_V$ and the full set of differential equations are defined in Appendix \ref{subsection:networkcentric}. It is interesting to note that $\theta_V(t)$ is conceptually related to its counterpart $\theta(s)$ (Eq. \eqref{eq:thetas}) of the finite-size algorithm, although they are obtained in a completely different manner. Figure \ref{fig:plbehaviourthreshold} displays  $\langle s \rangle_{\text{Volz}}$ as a function of $T$ and shows a remarkable improvement over the infinite size results. The remaining discrepancy with the simulations and $\langle s \rangle_N$ can be attributed to the mean field approximation used where only \emph{average} values are evolved in time.
\subsubsection{Moreno \emph{et al.} (2004): A compartmental approach}
A special class of models for the dynamical propagation on networks is the use of compartmental models where nodes of different degree are placed into different compartments \cite{pastor-satorras01_prl,pastor-satorras01_pre,moreno02_epjb}. We have integrated the SIR-like model of \cite{moreno02_epjb}, whose description is detailed in Appendix \ref{subsection:compartmental}, to construct the asymptotic average for a population $N$
\begin{equation}
\langle s \rangle_{\text{c}} = N \, \sum_k R_k( \infty ) \quad 
\end{equation}
where $R_k$ is the fraction of nodes that are recovered and of degree $k$. A slight correction to this model has been proposed in \cite{pastorsatorrasinternet04} and Appendix \ref{subsection:compartmental} explains how to obtain the corresponding asymptotic average $\langle s \rangle_{\text{c}}^{PV}$. The results seen in Fig. \ref{fig:plbehaviourthreshold} show large deviations of both models long before the critical $T_c$ for the network and appears valid only for small transmissibility. The source of failure in reproducing the simulations is quite subtle and has led us to propose an improved version.
\subsubsection{An improved compartmental approach}
Although the system \eqref{equation:Moreno} appears to be a natural choice, it neglects some important points. Firstly, all infectious nodes (except the first one) become infected from one of their neighbors. Since a link between two nodes will always remain attached to these two nodes, the ``effective'' degree of an infectious node is (at least) one less than its actual degree. We introduce this in the model by transferring nodes from $S_k$ to $I_{k-1}$ (instead of $I_k$). Secondly, an infectious node that infects a previously susceptible node along a given link cannot infect again along that same link; we should reduce the degree of infectious nodes, by one, each time they cause new infections. Thirdly, the same procedure must be done for infectious nodes linked to infectious or removed nodes (and applied to these infectious or removed nodes, as well). To put it differently, the knowledge that no infection occurs carries as much information as the knowledge that an infection occurs and both of these scenarios must be taken into account when modeling the dynamics of the system.

We have implemented these improvements in the original differential equations (see Appendix \ref{subsection:improvedcompartmental}) and the asymptotic result, say $\langle s \rangle_{\text{c}}^*$, is plotted in Fig. \ref{fig:plbehaviourthreshold}. The agreement is much better and essentially reproduces the result of Volz, $\langle s \rangle_{\text{Volz}}$. It seems that our modifications have 
nicely included the better part of the correct dynamics.

In contrast to the finite-size formalism that takes into account the full stochastic nature of the process, these three systems assumes that the ``average evolution'' occurs at each time. Such an approach seems to be more suitable for an endemic disease in a susceptible-infectious-susceptible (SIS) dynamics since the stochasticity is averaged when many nodes are infectious at the same time. When only a small amount of infectious nodes are present at some point in the evolution of the disease (\emph{e.g.} the very beginning), the stochastic nature of the dynamics can have tremendous impacts on the long-term behavior of the system. In particular, it is those very stochastic effects that allow for small components even beyond the epidemic threshold. Systems solely tracking averages cannot take into account this important behavior.
\subsection{Effective reproduction number}
\begin{figure*}[htb]
\includegraphics[width = .95\linewidth]{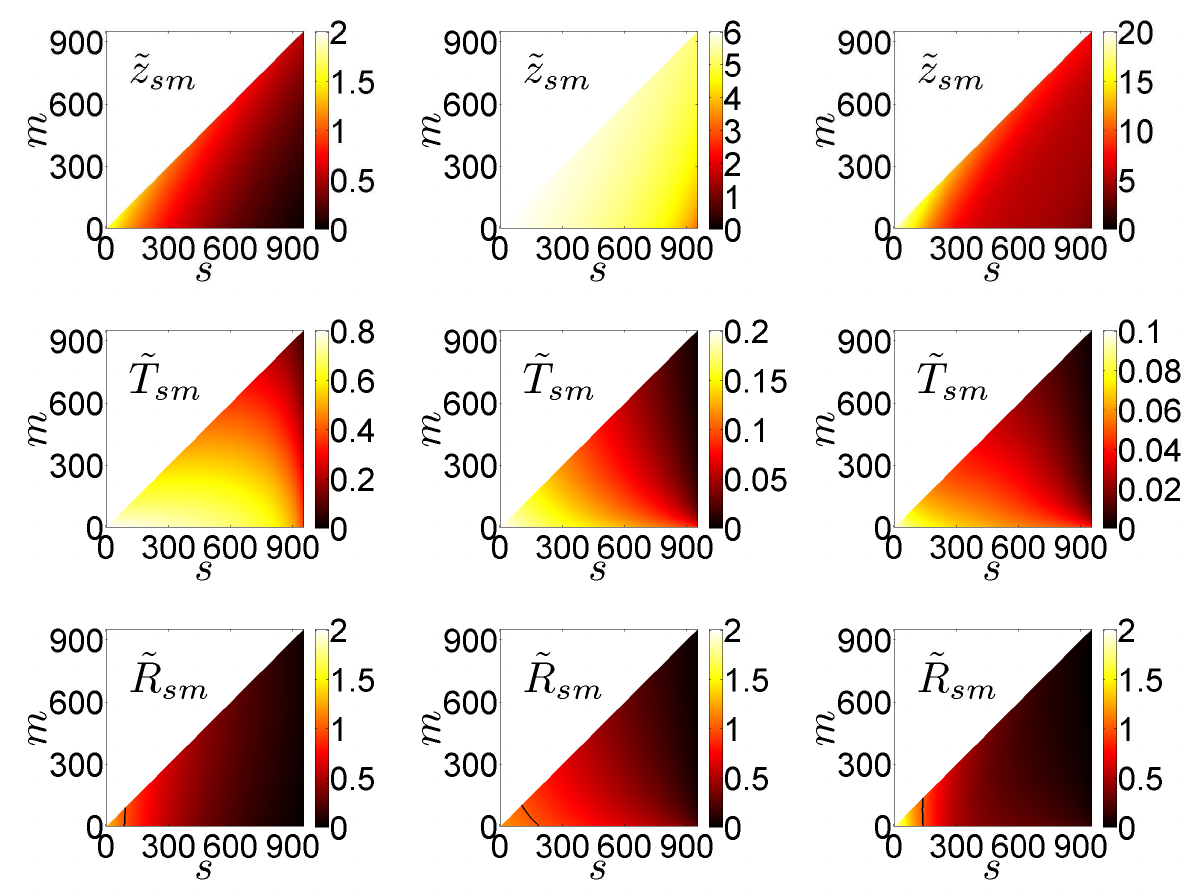}
\caption{(Color online) Effective reproduction number interpretation. For the 3 networks presented in Fig. \ref{fig:evodegdist} we show the expected excess degree of the infectious $\tilde{z}_{sm}$ , the effective transmissibility $\tilde{T}_{sm}$ and the corresponding effective reproduction number $\tilde{R}_{sm}$ for each ($s,m$) state. 
Th left, center, and right columns display the results of the power-law, the binomial, and the bimodal distributions respectively.
The solid black line in the $\tilde{R}_{sm} = \tilde{T}_{sm} \tilde{z}_{sm}$ plots corresponds to the ``threshold'' value $\tilde{R}_{sm}= 1$. \label{fig:fieldsplot}}
\end{figure*}
In order to establish a link between the present formalism and the classical epidemiological models, it is worth revisiting the interpretation of the basic reproductive number --- a key parameter in classical epidemiology \cite{andersonmay91,hethcote00_siamrev}. We can derive the mean excess degree of the infectious nodes as , $\tilde{z}_{sm} = \tilde{G}_{g - 1}(1;s,m)$, and together with the effective transmissibility $\tilde{T}_{sm}$, obtain the corresponding \emph{effective reproduction number} $\tilde{R}_{sm} = \tilde{T}_{sm} \tilde{z}_{sm}$ for each $(s,m)$ state. Figure \ref{fig:fieldsplot} shows the dependency of $\tilde{z}_{sm}$, $\tilde{T}_{sm}$, and $\tilde{R}_{sm}$ on $s$ and $m$ for the networks introduced in Fig. \ref{fig:evodegdist}. The behaviour of the power law distribution is dominated by the variability of $\tilde{z}_{sm}$, as $\tilde{T}_{sm}$ remains relatively uniform in the vicinity of the epidemic threshold. However, the converse is true for the binomial distribution; it is the variability of $\tilde{T}_{sm}$ that is responsible for the behavior of this distribution. The bimodal distribution is seen as a mixture of these two behaviors.
\subsection{Clustering and continuous time evolution}
Although most real-world contact networks show a clustered structure, and ways to incorporate this property into the dynamics are rapidly developing \cite{eames08_tpb,serrano06_prl, serrano06a_pre, miller08_arXiv}, our model does not at present explicitly account for clustering. In clustered networks, two neighbors of the same node $i$ are more likely to be neighbors of one another than of any other random node. In fact these ``triangles'' (or short loops) are not forbidden in the network ensemble introduced in Sec. \ref{section:InfiniteNetFormalism},  and over which our formalism is defined, but their number, however, is ruled  by randomness alone and is less than what would be expected in a typical realistic human population for instance. Hence, since it has recently been shown that clustering tends to decrease the number of infections in epidemic dynamics \cite{miller08_arXiv}, our model represents a worst case scenario for networks whose clustering properties are stronger than their random values. One promising way to improve upon the present framework, while keeping the generality of a \emph{pgf}-based formalism and allowing for better treatment of the evolution on clustered networks, would be to generalize the model to a bipartite graph structure where nodes (one part of the network) are assigned to different groups (the other part of the network) much in the same spirit as originally studied in \cite{newman03b_pre}.

Another aspect that needs further development is the relationship between discrete (generational) and  continuous time evolution. For systems where the underlying epidemiological dynamics is a discrete process in time (\emph{e.g.}, there is a constant time interval $\tau$ between successive infections), the generational formalism developed in Sec. \ref{section:InfiniteNetFormalism} directly represent the evolution of the system over time (infection of the $g$-th generation occurs at time $g \tau$). However, in most natural situations, the underlying dynamics is a continuous process. We are actively pursuing the issue of a general continuous epidemiological dynamics; the analysis will be the subject of a forthcoming contribution.
\section{Conclusion \label{section:conclusion}}
The emergence and re-emergence of infectious diseases pose a great threat to public health. The potential spread of a new pandemic strain of influenza or other emerging infection, such as SARS, may have a devastating impact on human lives and economies. There is an urgent need to develop reliable quantitative tools that can be used to compare the impact of various intervention strategies in real time. These tools must be able to incorporate the detailed structure of contact networks responsible for disease spread, as well as compare various intervention outcomes during the time of crisis, in a relatively short time span. In addition, these tools should be as equally applicable to large-scale networks as to finite-size networks, seeing that many interventions must be implemented not only globally, but locally (\emph{e.g.}, hospital settings, schools) as well.

In this paper, we have introduced and validated a theoretical framework that enables us to incorporate these two important aspects of disease outbreaks/epidemics, simultaneously. Specifically an extension to the existing formalism has been derived while keeping its appealing structure in terms of generating functions. With the introduction of the concept of generations and phase-space representation, Eq. \eqref{eq:Psi0gInf} --- for the infinite-size network with fixed transmissibility $T$ --- has been replaced by Eq. \eqref{eq:Psi0gFin} to account for finite-size effects through a modified generating function and an effective transmissibility. This is our finite-size, discrete time algorithm. One of its important features is its
ability to follow the complete diversity of outcomes, not only averages, from small clusters (outbreaks)
to the giant component (epidemic) within each generation.
A complete formalism including both finite-size and continuous time with or without correlations 
between transmissions is still lacking and its derivation is part of ongoing research.
\begin{acknowledgments}
BP would like to acknowledge the support of the Canadian Institutes of Health Research (grants no. MOP-81273 and PPR-79231), the Michael Smith Foundation for Health Research (Senior Scholar Funds) and the British Columbia Ministry of Health (Pandemic Preparedness Modeling Project). PAN and BD were supported by the above grants and PAN is also thankful to CIHR for a doctoral Scholarship. LJD is grateful to NSERC (Canada) and FQRNT (Qu\'ebec) for continuing support.
\end{acknowledgments}
\appendix
\section{The ratio $\rho_{s'm'} = \bigl\langle \langle \tilde{m} \rangle / m \bigr\rangle_{s'm'}$ 
%(Eq. \eqref{eq:rhoratiodef}) 
(Eq. 31) \label{section:rhosm}}
It is important to take into account that the number $\tilde{m}$ of new infections that actually occur in a finite network will typically be smaller than the number $m$ of new infections that would have happened in an infinite network. As presented in Sec. \ref{subsection:AddLossTrans}, the $m$ links leaving infectious nodes and causing new infections in the infinite network can lead to either susceptible nodes, other infectious nodes or recovered nodes. Of these links, only those leading to susceptibles nodes can actually result in new infections in the finite network and, moreover, a susceptible node targeted by more than one of these links contributes to only one transmission.

We note by $n_S$, $n_I$ and $n_R$ the number of links that are \emph{not forbidden} to join infectious nodes to susceptible nodes, other infectious nodes and recovered nodes respectively. Since there is no special restriction forbidding susceptible nodes to be linked to infectious ones, $n_S$ is simply the sum of the degrees of all susceptibles. In the same way, $n_I$ is the sum of the excess degree of the infectious nodes (since the sole restriction is that, for generations other than zero, there is at least one link from each infectious nodes to a recovered one). When the state of the infection is characterized by $s'$ and $m'$, these considerations translate to the mean values
\begin{align}
\langle n_S \rangle_{s'm'} & = (N - s')G_0^S{}'(1;s')       \label{eq:nSspmp} \\
\langle n_I \rangle_{s'm'} & = m' \tilde{G}_{g-1}'(1;s',m') \label{eq:nIspmp} \\
\langle m \rangle_{s'm'}   & = T \langle n_I \rangle_{s'm'} \label{eq:mspmp}  \quad ,
\end{align}
where $G_0^S(x;s')$ and $\tilde{G}_{g-1}(x;s',m')$ are defined in Eq. \eqref{eq:G0Sxs} and Eq. \eqref{eq:Gtildeg} respectively. More complicated constraints apply to $n_R$ and its mean value $\langle n_R \rangle_{s'm'}$ is obtained separately in Appendix \ref{section:nRsm}.

We develop two methods to evaluate $\rho_{s'm'} = \bigl\langle \langle \tilde{m} \rangle / m \bigr\rangle_{s'm'}$, the ratio of the mean number of new infections occurring in the finite network to the number of new infections that would have occurred in an infinite one. Both methods are based on the assumption that the $m$ links leading to infections in the infinite limit have the same \emph{a priori} probability of targeting any of the $n_S$, $n_I$ and $n_R$ targets, which is justified when $n_S + n_I + n_R$ is large compared to the degree of the node of highest degree in the network and is typically satisfied unless most of the network has been infected.

The first of these methods is quick and simple. The second method is based on a differential equation approach similar to the one used in Sec. \ref{subsection:evodegdist}. As both methods bring the same result, this second approach is an additional justification of the differential equation method for cases where an alternative approach is not known, as in Appendix \ref{section:nRsm} for instance.
\subsection{Direct approach}
Since the $m$ links causing infections in the infinite network can lead to $n_S + n_I + n_R$ potential targets in the finite one, each link belonging to a susceptible node (\emph{i.e.} one of the $n_S$ targets) has a probability
\begin{align}
\lambda & = \frac{m}{n_S + n_I + n_R} \label{eq:lambdadef}
\end{align}
to be one of these links. If that node is of degree $k$, there is a probability $1 - (1-\lambda)^k$ that \emph{at least one} of the $m$ links leads to it, causing a new infection. Using the degree distribution of the susceptibles obtained in Sec. \ref{subsection:evodegdist} and the fact that $N - s'$ susceptibles are left when the state of the infection is characterized by $s'$ and $m'$, the expected number of new infections in the finite network is then
\begin{align}
\langle \tilde{m} \rangle & = (N - s') \sum_k p_k^S(s') \left[ 1 - (1 - \lambda)^k \right] \nonumber \\
& = (N - s') \left[ 1 - G_0^S(1 - \lambda; s') \right] \quad . \label{eq:meantildemquickeasy}
\end{align}

Therefore, with the use of Eqs. (\ref{eq:nSspmp}--\ref{eq:meantildemquickeasy}), the ratio $\rho_{s'm'} = \bigl\langle \langle \tilde{m} \rangle / m \bigr\rangle_{s'm'}$ is provided by
\begin{align}
\rho_{s'm'} & = \frac{1}{T} \frac{(N - s')}{m'} \frac{\left[ 1 - G_0^S(1 - \lambda_{s'm'}; s') \right]}{\tilde{G}_{g-1}'(1;s',m')} \label{eq:rhospmpexpression}
\end{align}
where
\begin{align}
\lambda_{s'm'} & = \frac{\langle m \rangle_{s'm'}}{\langle n_S \rangle_{s'm'} + \langle n_I \rangle_{s'm'} + \langle n_R \rangle_{s'm'}} \quad , \label{eq:lambdaspmp}
\end{align}
and $\langle n_R \rangle_{s'm'}$ is given in Appendix \ref{section:nRsm}.
\subsection{Continuous approach \label{subsection:meanmtildecontinuous}}
In this approach, we relax the numerous constraints of the problem by changing the task to an ``assignment process'' evolving in time. By analogy, this process is similar to the system children use for their casual sports team when two leaders choose their team members one by one. It should be clear that the time dependency introduced in this manner is an artificial construct and that only the result at $t = \infty$ (``assignments of teams is completed'') is relevant.

We define $\alpha(t)$ as the pool of potential infections that occur in the infinite network and for which a target has yet to be assigned at time $t$. This definition implies the initial condition $\alpha(0) = m$; we must also have $\alpha(\infty) = 0$ (the process must assign a target to each potential infection). We also define $\beta(t)$, the pool of targets for which assignment of one of the potential infections does not lead to a new infection, yielding the initial condition $\beta(0) = n_R + (n_I - m)$. We finally define $\gamma_k(t)$, the pool of targets belonging to susceptible nodes of degree $k$ and to whom the assignment of a potential infection leads to an actual new infection in the finite network. From mean value considerations,
\begin{align}
\gamma_k(0) & = (N - s')\ k \ p_k^S(s')
\end{align}
with $\sum_k \gamma_k(0) = n_S$.

Since the time dependence is arbitrary, we assign targets to the potential infections at a rate proportional to their current population $\alpha(t)$, which is a simple way to obtain $\alpha(\infty) = 0$. With the definition
\begin{align}
\omega(t) & = \alpha(t) + \beta(t) + \sum_k \gamma_k(t) \quad ,
\end{align}
the probability of assigning a potential infection to other potential infections, other targets not leading to infections, or to targets leading to infections belonging to degree $k$ is $\alpha(t)/\omega(t)$, $\beta(t)/\omega(t)$ and $\gamma_k(t)/\omega(t)$, respectively. Moreover, upon assignment of a potential infection to a target belonging to $\gamma_k(t)$, we transfer the remaining $k - 1$ targets to $\beta(t)$ to ensure that later assignment to these targets will not lead to new infections. These considerations translate into a set of coupled nonlinear differential equations:
\begin{align}
\begin{split}
\frac{d\alpha(t)}{dt} & = -\alpha(t) \left[ 1 + \frac{\alpha(t)}{\omega(t)} \right] \\
\frac{d\beta(t)}{dt} & = -\alpha(t) \frac{\beta(t)}{\omega(t)} + \sum_k \alpha(t) \frac{\gamma_k(t)}{\omega(t)} (k - 1) \\
\frac{d\gamma_k(t)}{dt} & = -\alpha(t) \frac{\gamma_k(t)}{\omega(t)} k
\end{split}
\end{align}
for all $k$. By summing these terms, we note that $d\omega(t)/{dt} = -2\alpha(t)$, together with $d\alpha(t)/dt$, yields the solution
\begin{align}
\frac{\alpha(t)}{\omega(t)} & = \frac{\lambda}{\lambda + (1 - \lambda)e^t}
\end{align}
where $\lambda = \alpha(0)/\omega(0) = n_T/(n_S + n_I + n_R)$. This result allows us to rewrite the equation governing $\gamma_k(t)$ as
\begin{align}
\frac{d\gamma_k(t)}{dt} & = \frac{-k \gamma_k(t)}{1 + (\lambda^{-1}-1)e^t} \quad ,
\end{align}
which is completely decoupled from the rest of the system. This equation has the solution
\begin{align}
\gamma_k(t) & = \gamma_k(0) \left( \lambda e^{-t} + 1 - \lambda \right)^k \quad ,
\end{align}
and produces the limit
\begin{align}
\gamma_k(\infty) & = \gamma_k(0)\left(1 - \lambda\right)^k \quad .
\end{align}
The number of nodes of degree $k$ that are infected in the process is thus
\begin{align}
\frac{\gamma_k(0) - \gamma_k(\infty)}{k} & = \frac{\gamma_k(0)}{k}\bigl( 1 - (1-\lambda)^k \bigr) \nonumber \\
& = (N - s') p_k^S(s') \bigl(1 - (1 - \lambda)^k \bigr) \quad .
\end{align}
Using the initial conditions, we obtain the average total number of infections in the finite network
\begin{align}
\langle \tilde{m} \rangle & = (N - s')\left( \sum_k p_k^S(s') - \sum_k p_k^S(s')(1-\lambda)^k \right) \nonumber \\
& = (N-s')\left[ 1 - G_0^S\left(1-\lambda;s'\right) \right] \quad .
\end{align}
Notice that this expression is the same as Eq. \eqref{eq:meantildemquickeasy} and thus leads to the same $\rho_{s'm'}$, \emph{i.e.} Eq. \eqref{eq:rhospmpexpression}.
\section{The average value $\langle n_R \rangle_{s'm'}$ \label{section:nRsm}}
While it is quite easy to obtain mean values or even distributions for $n_I$, $n_S$ and $m$, an independent method is required to evaluate the quantity $n_R$. Here, we use a continuous approach similar to that of Appendix \ref{section:rhosm}.

Again, we design a differential equation governing the evolution of the continuous counterpart of $n_R$, denoted $\eta(s)$, as a function of the number of infections, $s$. If the $k-1$ excess degrees belonging to a newly infected node, of degree $k$, are linked to susceptible nodes, we expect a fraction $1-T$ of them to be unsuccessful in transmitting the infection to others and thus, contribute to $\eta(s)$. However, we expect a fraction, $\eta(s)/\langle n_S(s) \rangle$, of the $k-1$ excess degrees to be linking to recovered nodes, with $\langle n_S(s) \rangle = (N-s)z_1^S(s)$. These last links actually \emph{reduce} the value of $\eta(s)$, because links between susceptible and recovered nodes are converted to links between recovered and infectious nodes. Defining
\begin{align}
z_2^S(s) &= \sum_k k(k-1) p_k^S(s) = G_0^S{}''(1;s) \quad ,
\end{align}
the differential equation considering these effects reads
\begin{align}
\frac{d \eta(s)}{ds} & = \sum_k \frac{k p_k^S(s)}{z_1^S(s)} (k-1) \nonumber \\
& \qquad \times \left[ \left( 1 - \frac{\eta(s)}{\langle n_S(s) \rangle} \right)(1-T) - \frac{\eta(s)}{\langle n_S(s) \rangle} \right] \nonumber \\
& = \frac{z_2^S(s)}{z_1^S(s)} \left[ (1-T) - (2-T)\frac{\eta(s)}{\langle n_S(s) \rangle} \right] \quad .
\end{align}
Together with the initial condition $\eta(1) = (1-T)z_1^S(1)= (1-T)z_1$, this differential equation can be integrated numerically to provide an expectation value for the number of links between recovered and susceptible nodes when the size of the outbreak/epidemic is $s$.

The quantity $\langle n_R \rangle_{s'm'}$, required to fix $\lambda_{s'm'}$ in Appendix \ref{section:rhosm}, is the mean number of links emerging from recovered nodes that are allowed to target infectious nodes when we know the current state of the epidemic is characterized by $s'$ and $m'$. As $m'$ is the current number of infectious nodes, the size of the outbreak/epidemic was $s'-m'$ when these infectious nodes were still susceptible nodes (in the previous generation). At that time, we expect $\eta(s'-m')$ links to join susceptible and recovered nodes; this number is equal to the expectation value $\langle n_R \rangle_{s'm'}$.
\section{Some dynamical models \label{section:appendixsomedynamicalmodels}}
The following models are all defined for the same network ensemble as the one presented in Sec. \ref{section:InfiniteNetFormalism}, \emph{i.e.} their degree distribution $\{p_k\}$ is generated by $G_0(x)$.
\subsection{Volz (2008): A network-centric approach \label{subsection:networkcentric}}
At time $t$, let $\theta_V(t)$ be the fraction of degree one nodes that remain susceptible, $p_I(t)$, the probability that a susceptible node be connected to an infectious one and $p_S(t)$, the probability that a susceptible node be connected to a susceptible one, then
% \begin{widetext}
% \begin{align}
% \begin{split}
% \frac{d\theta_V(t)}{dt} & = -r p_I(t) \theta_V(t) \\
% \frac{dp_I(t)}{dt}      & = r p_S(t) p_I(t) \theta_V(t) \frac{G_0''\bigl( \theta_V(t) \bigr)}{G_0'\bigl( \theta_V(t) \bigr)} - r p_I(t) \bigl( 1 - p_I(t) \bigr) - \mu p_I(t) \\
% \frac{dp_S(t)}{dt}      & = r p_S(t) p_I(t) \left( 1 - \theta_V(t) \frac{G_0''\bigl( \theta_V(t) \bigr)}{G_0'\bigl( \theta_V(t) \bigr)} \right)
% \end{split} \quad ,
% \end{align}
% \end{widetext}
\begin{align}
\begin{split}
\frac{d\theta_V(t)}{dt} & = -r p_I(t) \theta_V(t) \\
\frac{dp_I(t)}{dt}      & = r p_S(t) p_I(t) \theta_V(t) \frac{G_0''\bigl( \theta_V(t) \bigr)}{G_0'\bigl( \theta_V(t) \bigr)} \\
& \qquad - r p_I(t) \bigl( 1 - p_I(t) \bigr) - \mu p_I(t) \\
\frac{dp_S(t)}{dt}      & = r p_S(t) p_I(t) \left( 1 - \theta_V(t) \frac{G_0''\bigl( \theta_V(t) \bigr)}{G_0'\bigl( \theta_V(t) \bigr)} \right)
\end{split} \quad ,
\end{align}
where $r$ is the force of infection, the constant rate at which infectious nodes infect a neighbor and $\mu$ the recovery rate at which infected nodes become recovered. For the purpose of calculations, we have chosen $\mu = 1$ and a related transmissibiliy $T = 1 - e^{-r}$ \cite{newman02_pre}. The initial conditions have been set to $\theta_V(0) = 1 - \epsilon$, $p_I(0) = \epsilon/(1-\epsilon)$ and $p_S(0) = (1 - 2\epsilon)/(1 - \epsilon)$ for $\epsilon$ satisfying $G_0\bigl( 1 - \epsilon \bigr) = 1 - 1/N$. The asymptotic ($t \to \infty$) average size of outbreak and/or epidemic is then given by
\begin{equation}
\langle s \rangle_{\text{Volz}} = N \, \left( 1 - G_0\bigl( \theta_V( \infty ) \bigr) \right) \quad .
\end{equation}
\subsection{Moreno \emph{et al.} (2004): A compartmental approach \label{subsection:compartmental}}
At time $t$, let $S_k(t)$ be the fraction of nodes that are susceptible ($I_k(t)$ and $R_k(t)$ for the infectious and the removed nodes respectively) \emph{and} of degree $k$ subject to the normalization
\begin{equation}
\sum_k \bigl( S_k(t) + I_k(t) + R_k(t) \bigr) = 1 \quad .
\end{equation}

With the probability that any given link  pointing to an infected node is
\begin{equation}
\Theta(t) = \frac{\displaystyle\sum_k k I_k(t)}{\displaystyle\sum_{k'} k' \bigl(S_{k'}(t) + I_{k'}(t) + R_{k'}(t)\bigr)} \quad , \label{eq:Theta}
\end{equation}
and $r$ defined as in Sec. \ref{subsection:networkcentric}, the dynamics obeys the differential equations
\begin{align}
\begin{split}
\frac{dS_k(t)}{dt} & = -r \, \Theta(t) \, k \, S_k(t) \\
\frac{dI_k(t)}{dt} & = -I_k(t)  + r \, \Theta(t) \, k \, S_k(t) \\
\frac{dR_k(t)}{dt} & = \phantom{+} I_k(t)
\end{split} \quad . \label{equation:Moreno}
\end{align}

Solving this system for the initial conditions $S_k(0) = (1 - 1/N) p_k$, $I_k(0) = p_k/N$ and $R_k(0) = 0$ leads to the asymptotic average size of outbreak and/or epidemic
\begin{equation}
\langle s \rangle_{\text{c}} = N \, \sum_k R_k( \infty ) \quad .
\end{equation}

Note that a slight correction to this model has been proposed in \cite{pastorsatorrasinternet04} in order to take into account that infectious nodes acquired their infection along one of their links and that new transmission is impossible along that link. The corresponding asymptotic average size of outbreak and/or epidemic $\langle s \rangle_{\text{c}}^{PV}$ is obtained in the same way as for the compartmental approach except that the factor $k$ in Eq. \eqref{eq:Theta} has to be replaced by $k - 1$.
\subsection{An improved compartmental approach \label{subsection:improvedcompartmental}}
Incorporating in Eq. \eqref{equation:Moreno} the improvements described in Sec.\ref{section:discussion} results in a new differential system
% \begin{widetext}
% \begin{align}
% \begin{split}
% \frac{dS_k^*(t)}{dt} & = -r \, \Theta^*(t) \, k \, S_k^*(t) \\
% \frac{dI_k^*(t)}{dt} & = -I_k^*(t) + r \, \Theta^*(t) \bigl(k + 1\bigr) S_{k + 1}^*(t) + r \bigl( \Theta^*(t) + 1 \bigr) \Bigl[ \bigl(k + 1\bigr) I_{k + 1}^*(t) - k I_k^*(t) \Bigr] \\
% \frac{dR_k^*(t)}{dt} & = \phantom{+}I_k^*(t) + r \, \Theta^*(t) \Bigl[ \bigl(k + 1\bigr) R_{k + 1}^*(t) - k R_k^*(t) \Bigr]
% \end{split} \quad . \label{equation:ImprovedMoreno}
% \end{align}
% \end{widetext}
\begin{align}
\begin{split}
\frac{dS_k^*(t)}{dt} & = -r \, \Theta^*(t) \, k \, S_k^*(t) \\
\frac{dI_k^*(t)}{dt} & = -I_k^*(t) + r \, \Theta^*(t) \bigl(k + 1\bigr) S_{k + 1}^*(t) \\
& \qquad + r \bigl( \Theta^*(t) + 1 \bigr) \Bigl[ \bigl(k + 1\bigr) I_{k + 1}^*(t) - k I_k^*(t) \Bigr] \\
\frac{dR_k^*(t)}{dt} & = \phantom{+}I_k^*(t) + r \, \Theta^*(t) \Bigl[ \bigl(k + 1\bigr) R_{k + 1}^*(t) - k R_k^*(t) \Bigr]
\end{split} \quad . \label{equation:ImprovedMoreno}
\end{align}
$\Theta^*(t)$ is the the corresponding expression for $\Theta(t)$ (Eq. \ref{eq:Theta}) and the integration is carried out under the same initial conditions. The asymptotic average size of outbreak and/or epidemic is then also
\begin{equation}
\langle s \rangle_{\text{c}}^* = N \, \sum_k R_k^*( \infty ) \quad .
\end{equation}
Both infinite differential systems \eqref{equation:Moreno} and \eqref{equation:ImprovedMoreno} are truncated at $k_\text{max} =  50$ in numerical simulations.
%
% \bibliographystyle{apsrev}
% \bibliography{../bib/networks}
%

%
\end{document}